\documentclass[10pt,journal,final]{IEEEtran}
\usepackage{amsmath,amsfonts}
\usepackage{amssymb}
\usepackage{amsthm}

\usepackage{xcolor}

\renewcommand{\qed}{\hfill \textcolor{black}{$\blacksquare$}}
\usepackage{flushend}
\usepackage{algorithm}
\usepackage{algpseudocode}
\usepackage{array}
\usepackage{stfloats}
\usepackage{url}
\usepackage{verbatim}
\usepackage{graphicx}
\usepackage{cite}
\usepackage{setspace}
\usepackage{booktabs}
\usepackage{multirow}
\usepackage{cuted}
\usepackage{subfigure}   
\usepackage{color}  
\usepackage{subeqnarray}
\usepackage{cases}

\begin{document}

\title{Secure Communication via Modulation \\ Order Confusion}

\author{Jingyi Wang and Fanggang Wang
        
\thanks{}
\thanks{}}

\maketitle

\begin{abstract}
With the advancement of wireless communications, secure transmission faces growing challenges. Modulation classification, essential for decoding intercepted signals, has enabled eavesdroppers to threaten communication security.
In this paper, we propose a secure communication approach based on modulation order confusion (MOC), which disguises the original signal as a higher-order or lower-order modulation to mislead the classifiers of Eve and prevent information leakage.
For the single-antenna system, we develop two schemes, i.e., the symbol-random-mapping scheme and the symbol-time-diversity scheme.
The former one disguises the low-order modulation as high-order modulation by probabilistically mapping each original symbol to multiple alternatives, with the mapping probability optimized through convex optimization to balance the error performance and security. 
The latter one disguises the high-order modulation as low-order modulation by encoding part of the information in varying symbol durations, thereby reducing the apparent modulation order without losing information. Both the two schemes require a customized receiver design at Bob for correct demodulation.
For the multi-antenna system, we propose the MOC schemes that require no receiver modifications. Two signal design strategies are introduced, namely a series-expansion-based scheme where a Taylor-series-based confusion transmitter decomposes the original signal into structured components, and a constellation-path-design confusion scheme where the signal is expressed as a sum of general components along the predefined trajectories.
We further extend the multi-antenna MOC approach to the reconfigurable intelligent surface (RIS) systems, presenting a joint design of the RIS reflection and the beamformer. Even if Eve performs perfect blind source separation, it cannot infer the true modulation format.
Numerical results demonstrate that the proposed MOC schemes can effectively resist existing modulation classifiers, including both the deep-learning-based and the expert-knowledge-based approaches, without compromising communication quality.
\end{abstract}


\section{Introduction}
Wireless communication has rapidly developed with the growth in the number of intelligent devices and the expansion of network coverage \cite{01nguyen2021security}. However, due to the broadcast nature of wireless communication medium, security in wireless communications faces severe challenges \cite{02zou2016survey,03angueira2022survey,04jameel2018comprehensive}. Common approaches to enhancing wireless communication security include encryption-based schemes and physical layer security (PLS) \cite{05bloch2008wireless,06goel2008guaranteeing}. The former type is typically computationally intensive, making it difficult to deploy in low-cost devices. In recent years, PLS schemes have gained attention due to their low complexity and high reliability.

Regarding the wireless communication systems, modulation classification is a crucial process for non-cooperative receivers before signal demodulation and decoding. Consider a communication system consisting of Alice, Bob, and Eve,\footnote{Building on the previous studies in wireless security, we use the terms “Alice”, “Bob”, and “Eve” to respectively denote the transmitter, the intended receiver, and the eavesdropper.} where Bob demodulates using a pre-agreed modulation format as prior knowledge, while Eve must identify the modulation format of Alice before recovering data. Therefore, the study of modulation classification and anti-classification of wireless signals has attracted widespread attention from researchers.

Research on modulation classification by scholars began as early as the last century.
Earlier work can be coarsely divided into two classes, i.e., the likelihood-based methods and the feature-based  methods, which are derived from statistical decision theory and pattern recognition theory respectively.
The performance of the likelihood-based methods is optimal, yet finding closed solutions is challenging. Moreover, the likelihood-based methods exhibit limited robustness in dynamic environments affected by issues like fading, frequency offset, and phase offset \cite{07huan1995likelihood,08yucek2004novel,09zhu2018likelihood,10wei2000maximum}. Conversely, the feature-based methods are simpler to implement and provide near-optimal performance, prompting extensive research by the scholars. 
Many features are extracted using the expert knowledge for the modulation classification, such as the cumulants \cite{11swami2000hierarchical,12wu2008novel,13chang2015cumulants}, the cyclostationary spectrum \cite{14dobre2004robust,15dobre2010cyclostationarity}, the wavelet transformation \cite{16park2008automatic}, and so on.
Recently, advancements in neural network architectures, enhancements in optimization algorithms, and updates to hardware equipment have propelled deep learning to the forefront of modulation classification. These developments have enabled deep learning to excel and progressively supplant traditional methods in this field \cite{17o2017introduction,VGGo2018over}.
Numerous neural networks specifically designed for modulation classification tasks have been developed, such as VGG \cite{VGGo2018over}, SCGNet \cite{SCGNettunze2020sparsely}, WSMF \cite{WSMFqi2020automatic}, ChainNet \cite{ChainNethuynh2020chain}, and CGDNet \cite{22njoku2021cgdnet}.
Deep-learning-based modulation classification methods offer a robust and adaptable approach to handling complex wireless signal environments. 
Current modulation classification strategies primarily cater to single-antenna systems. Recognizing modulations in multi-antenna systems poses greater challenges due to each antenna receiving mixed signals from various transmit antennas. Classic multi-antenna modulation classification methods include those based on independent component analysis \cite{24choqueuse2009blind} and expectation maximization \cite{25zhu2014blind}, both aimed at first separating the multipath signals before individually identifying their modulations. In contrast, methods utilizing moment statistics have been introduced to capture the comprehensive characteristics of multi-antenna signals for modulation recognition \cite{26liu2017blind,27hassan2011blind}. 
Modulation classification schemes have become highly advanced, enabling Eve to easily decrypt the transmit signals of Alice, which poses a significant security risk. Consequently, anti-modulation-classification techniques have gained increasing attention.

Nowadays, research into communication security focused on modulation methods is still in its early stages.
In \cite{28althunibat2017physical}, the authors introduce a phase-based adaptive modulation scheme to enhance communication security. This approach dynamically adjusts the modulation type based on the channel phase, but requires Alice and Bob to perform accurate channel estimation; otherwise, it will lead to a decline in the Bit Error Rate (BER). 
The researchers in \cite{29bang2020secure} introduce a security modulation technique based on constellation mapping confusion, which randomly alters the constellation mapping rules for each data packet, instead of using a predefined mapping for each packet. This scheme requires Alice and Bob to share the mapping rules, and it provides better security performance only under high-order modulation.
Recently, adversarial machine learning has received significant attention for its potential to induce subtle perturbations that mislead deep learning models. This approach has been adapted to secure modulation classification systems in wireless communication, thwarting attempts by eavesdroppers to correctly classify signals. Research categorizes these adversarial tactics into two main types: independent attack and add-on attack. Independent attack involve external jammers introducing perturbations without altering the original signal, whereas add-on attacks integrate perturbations directly into the signal by the transmitter.
The authors in \cite{34sadeghi2018adversarial} introduce the white-box and the black-box attacks targeting a deep-learning-based modulation classifier, showing that these methods require less transmit power for successful misclassification than traditional jamming techniques. Research documented in \cite{35lin2020adversarial} assesses various attack strategies on this classifier, revealing that adversarial perturbations significantly reduce classification accuracy, with iterative methods outperforming one-step approaches in efficacy. Furthermore, \cite{36kim2021channel} illustrates that adversarial perturbations are ineffective in scenarios where fading channels aren't accounted for in the perturbation design.
In \cite{37he2023anti}, the authors provide a general framework for optimizing additive attacks, effectively integrating factors that impact both communication performance and security.
These anti-modulation-classification strategies based on adversarial attacks can effectively suppress the performance of the deep-learning-based modulation classifiers. However, due to the presence of artificial perturbations, the BER of Bob will inevitably be affected. Moreover, while these subtle disturbances can curb the efficacy of deep-learning-based modulation classification algorithms, their ability to suppress the classifiers based on expert features is limited.

The main contributions of this paper are:
\begin{itemize}
\item{
Propose a secure communication approach based on modulation order confusion (MOC), which ensures transmission confidentiality by preventing Eve from accurately identifying the modulation format used on the legitimate link. 
We design MOC schemes for both single-antenna and multi-antenna systems. The former requires a custom receiver design, while the latter does not.
Compared to the existing secure communication and anti-modulation-classification schemes, the proposed schemes do not rely on any knowledge of Eve. Users can flexibly disguise the original signal as a higher-order or lower-order modulation to balance security and error performance. 
}
\item{ 
In the single-antenna system, we propose two MOC schemes for disguising the original signal as a higher-order or lower-order modulation. The first scheme, based on symbol random mapping, probabilistically maps each original symbol to higher-order modulation symbols. The mapping probability is optimized to balance the BER and security performance. The second scheme, based on symbol time diversity, represents each original symbol using a reduced number of lower-order symbols and encodes part of the information in varying symbol durations. At the receiver, we design the corresponding demodulators, that is, the symbol-reverse-mapping decoder for the first scheme and the dynamic-programming-based decoder for the second one. 
}
\item{ 
In the multi-antenna system, we also propose two MOC schemes: one based on the series expansion and the other on the constellation path design. For the former, we develop the Taylor-series-based confusion scheme, which decomposes the original signal into multiple specially structured components. For the latter, the signal is represented as a sum of the generalized components following the predefined constellation trajectories. 
Then, we extend the multi-antenna MOC approach to the reconfigurable intelligent
surface (RIS) systems, and present a joint design of the RIS reflection and the beamformer.
All these confusion schemes do not require any receiver design. 
}
\end{itemize}

The remainder of this paper is organized as follows.
Section II introduces the MOC schemes in the single-antenna system.
The MOC schemes in the multi-antenna system is proposed in Section III. 
In Section IV, we present the extension of the proposed scheme to the RIS-assisted systems.
Simulation results are provided in Section V. 
Section VI concludes the paper.

\textit{Notation:} 
Variables, vectors, and matrices are written as italic letters $x$, bold italic letters $\boldsymbol{x}$, and bold capital italic letters $\boldsymbol{X}$, respectively. For any vector $\boldsymbol{x}$, ${\text{diag}}\left\{ \boldsymbol{x} \right\}$ denotes a diagonal square matrix whose diagonal consists of the elements of $\boldsymbol{x}$. For any square matrix $\boldsymbol{X}$, ${[\boldsymbol{X}]_{{\text{diag}}}}$ denotes a diagonal square
matrix formed by the diagonal elements of $\boldsymbol{X}$. ${\mathbb{C}^M}$ denotes the set of all $M$ complex-valued vectors and ${{\mathbb{C}}^{M \times N}}$ denotes the set of all ${M \times N}$ complex-valued matrices.
$\Re \{  \cdot \} $ represents the real part of the element.
The operator  ${\left(  \cdot  \right)^\mathsf{T}}$ denotes the transpose.
Define ${{\mathcal{I}}_N} = \left\{ {1,2, \ldots ,N} \right\}$ as a shorthand as the index set.
The imaginary unit of a complex number is denoted by $\jmath = \sqrt {-1} $.
The default base of the logarithm is $2$.

\section{Modulation Order Confusion for Single-Antenna Communication}
In this section, we first introduce a single-antenna communication system. 
Two MOC schemes are then provided to confuse the modulation order to appear as a higher or lower order. Finally, we analyze and compare these two schemes.

\subsection{Single-Antenna System Model}

\begin{figure}[t]
\centering
\includegraphics[width=8.5cm]{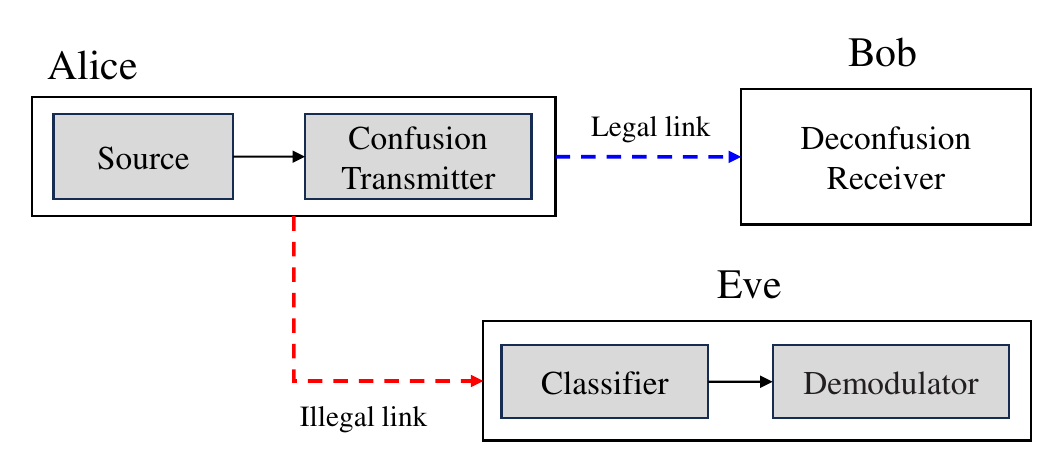}
\caption{Illustration of the single-antenna system.}
\label{systemmodel1}
\end{figure}

We consider a wireless communication system where Alice communicates with Bob, and Eve attempts to identify the modulation format of Alice, as illustrated in Figure \ref{systemmodel1}.
Alice, Bob, and Eve are each equipped with a single antenna. 
To prevent Eve from easily identifying the modulation format, Alice employs a confusion transmitter to disguise the signals as another modulation format.
Specifically, the original secure signal of Alice is ${\boldsymbol{s}} = {\big[{s_1},{s_2}, \ldots ,{s_L}\big]^\mathsf{T}}$, and the transmit signal after confusion is denoted as 
${\boldsymbol{x}} = {\big[{x_1},{x_2}, \ldots ,{x_{L}}\big]^\mathsf{T}}$ where 
${s_l} \in {\mathcal{A}},~{x_{l}} \in {\mathcal{B}},~l \in {{\mathcal{I}}_{L}}$, 
${\mathcal{A}} = {\{{a_1},{a_2}, \ldots ,{a_M}\}}$ and ${\mathcal{B}} = {\{{b_1},{b_2}, \ldots ,{b_N}\}}$ are the alphabets of the modulation formats before and after confusion, respectively; 
$M$ and $N$ denote the modulation orders of $\mathcal{A}$ and $\mathcal{B}$, respectively, with $M \ne N$;
$L$ is the length of the signal sequences.

The received signal sequence of Bob is ${\boldsymbol{y}} = {\big[y_1,y_2, \ldots ,y_{L}\big]^\mathsf{T}}$, and the $l$-th symbol of $\boldsymbol{y}$ is 
\begin{equation}
   y_{l} = hx_{l} + \mu,\quad {l} \in {{\mathcal{I}}_{L}}
\end{equation}
where $h$ is the channel state information (CSI) between Alice and Bob,
$\mu$ denotes the additive Gaussian white noise (AWGN) at Bob following from a
circularly symmetric complex Gaussian distribution, i.e., $\mu \sim {\mathcal{C}\mathcal{N}}(0,{\sigma ^2})$. 
Eve receives the signal ${\boldsymbol{z}} = {[{z_1},{z_2}, \ldots ,{z_{L}}]^\mathsf{T}}$ from Alice, and the $l$-th symbol of $\boldsymbol{z}$ is given by
\begin{equation}
   z_{l} = gx_{l} + \nu,\quad {l} \in {{\mathcal{I}}_{L}}
\end{equation}
where $g$ is the CSI of the Alice-Eve link, $\nu$ is the AWGN at Eve, i.e., $\nu \sim {\mathcal{C}\mathcal{N}}(0,{\gamma  ^2})$. 

\emph{Remark 1:}
In the single-antenna system, Bob is aware of the confusion strategy and employs a specially designed receiver to effectively deconfuse the received signal.
Eve is unavailable to decode due to the incorrect modulation recognition, which will be elaborated in the subsequent subsections. 
In the scenarios where complexity and cost considerations discourage redesigning the receiver of Bob to align with the confusion transmitter of Alice, we propose a multi-antenna secure communication system, which will be introduced in the next section.

\subsection{Low-to-High-Order Confusion}
In this section, we first introduce a low-to-high-order confusion transmitter based on symbol random mapping (SRM). Then, we formulate a mapping probability optimization problem to balance security and error performance. Finally, the design of the receiver of Bob is presented.
\subsubsection{Confusion Transmitter}
The concept of SRM involves mapping each symbol from the original modulation format $\mathcal{A}$ to one of several symbols in the modulation format $\mathcal{B}$ after confusion, using specific probabilities. This approach is designed to make Eve overestimate the modulation order of the legitimate signals.
In this low-to-high-order confusion scenario, we have $M<N$.
Initially, the alphabet of the modulation format $\mathcal{B}$, with an order $N$, is divided into $M$ ordered subsets, denoted as 
\begin{equation}
{{\mathcal{B}}_m} = \big\{ b_1^{(m)},b_2^{(m)}, \ldots ,b_{{K_m}}^{(m)}\big\} ,\quad m \in {{\mathcal{I}}_M}
\end{equation}
satisfying 
\begin{equation}
   \bigcap\limits_{m \in {{\mathcal{I}}_{M}}} {{{\mathcal{B}}_m}}  = \O,~\bigcup\limits_{m \in {{\mathcal{I}}_{M}}} {{{\mathcal{B}}_m}}  = {\mathcal{B}}
\end{equation}
where $K_m,~m\in {{\mathcal{I}}_{M}}$ denotes the number of elements in $\mathcal{B}_m,~m\in {{\mathcal{I}}_{M}}$.
Then, each symbol ${a_m}$ in $\mathcal{A}$ is mapped to a symbol in $\mathcal{B}_m$ with a probability vector ${{\boldsymbol{p}}_m}$, given by
\begin{equation}
{{\boldsymbol{p}}_m} = \big[p_1^{(m)},p_2^{(m)}, \ldots ,p_{{K_m}}^{(m)}\big],\quad m \in {{\mathcal{I}}_M}
\end{equation}
where $p_i^{(m)},~i \in {{\mathcal{I}}_{{K_m}}}$ denotes the probability that the symbol ${a_m}$ is mapped to $b_i^{(m)},~i \in {{\mathcal{I}}_{{K_m}}}$.
In practice, since both $M$ and $N$ are powers of $2$, without loss of generality, we evenly divide $\mathcal{B}$ into $M$ groups, each containing $K = \frac{N}{M}$ elements.
When the symbols in $\mathcal{B}$ are evenly divided into $M$ groups and the probabilities in $\boldsymbol{p}_m,~m \in {{\mathcal{I}}_M}$ are equal, the probabilities of $b_n,~n\in {{\mathcal{I}}_{{N}}}$ in the confused signal $\boldsymbol{x}$ are equal and independent, and thus, the physical characteristics of $\boldsymbol{x}$ fully conform to the modulation format $\mathcal{B}$.
However, Alice typically opts for lower-order modulation $\mathcal{A}$ due to poor channel conditions and noise. Fully disguising it as the higher-order modulation $\mathcal{B}$ can enhance security but significantly compromise the error performance. Therefore, in the next subsection, we design an optimization scheme to balance the security and error performance.

\subsubsection{Mapping Probability Optimization Problem}
We use the BER $\vartheta $ as the metric for the error performance, while the Kullback–Leibler (KL) divergence between the symbol probability distributions before and after confusion serves as the metric of the security performance.
The goal of Alice is to minimize the KL divergence by optimizing the probability vectors ${{\boldsymbol{p}}_m},~m \in {{\mathcal{I}}_M}$ subject to the BER constraint.
The specific optimization problem is presented in the following lemma.

\emph{Lemma 1:}
In the BER constraint, the problem of minimizing the KL divergence between the original and the confused symbol distributions can be formulated as
\begin{subequations}
  \label{Problem1}
  \begin{align} 
  \label{Problem1a}
{\mathcal{P}1}:~\mathop {{\text{min}}}\limits_{{{{\boldsymbol{p}}_m}},m \in {{\mathcal{I}}_M}} \hspace{1em}&\log K - \frac{1}{M}\sum\limits_{m = 1}^M {H({{\boldsymbol{p}}_m})}\\
\label{Problem1b}{\text{s.t.}}\hspace{2.2em}&\vartheta  \leqslant \eta \\
\label{Problem1c}&\sum\limits_{i = 1}^{{K}} {p_i^{(m)}}  = 1,\quad m \in {{\mathcal{I}}_M}
  \end{align}
\end{subequations}
where $H({{\boldsymbol{p}}_m}) =  - \sum\limits_{i = 1}^K {p_i^{(m)}\log p_i^{(m)}} $, $\eta $ is the BER constraint given by
\begin{align}
\vartheta  = \frac{1}{{M\log M}}\sum\limits_{m = 1}^M {\sum\limits_{i = 1}^{{K}} {p_i^{(m)}\sum\limits_{j = 1}^{N - {K}} {\rho _{i,j}^{(m)}Q\left( {\frac{{{\Delta _{i,j}^{(m)}}}}{{\sqrt 2 \sigma }}} \right)} } } .
\end{align}

\emph{Proof:}
The proof is given in Appendix.\qed

The problem $\mathcal{P}1$ is a convex entropy minimization problem and can be directly solved using CVX \cite{grant2009cvx}.

\subsubsection{Deconfusion Receiver}
The receiver of Bob can be designed as a cascade of a conventional demodulator and a demapper. Initially, Bob demodulates $\boldsymbol{y}$ to obtain the symbols $\boldsymbol{x}$ of the modulation format $\mathcal{B}$.
Then, based on Alice's grouping of the symbols in $\mathcal{B}$, Bob remaps each symbol in $\boldsymbol{x}$ back to the corresponding symbols in $\mathcal{A}$ to retrieve $ s $, that is,
\begin{equation}
{s_l} = \sum\limits_{m \in {\mathcal{I}_M}} {{a_m}{{\mathbb{I}}_{\{ {x_l} \in {\mathcal{B}_m}\} }}} ,\quad l \in {\mathcal{I}_L}
\end{equation}
where ${{{\mathbb{I}}_{\{ {x_l} \in {\mathcal{B}_m}\} }}}$ is an indicator function that equals $1$ if ${{x_l} \in {\mathcal{B}_m}}$ and $0$ otherwise.
This process is difficult for Eve to implement for the following reasons. First, Eve demodulates $\boldsymbol{z}$ based on the results of modulation classification, which makes it hard to detect that the signals have been mapped. Second, even if Eve realizes that the transmit symbols have been mapped, it lacks any information to determine the mapping rules. Third, Alice can dynamically adjust the grouping rules and mapping probabilities during communication with Bob.

\subsection{High-to-Low-Order Confusion}
In this section, we propose a scheme based on symbol time diversity (STD) to confuse higher-order modulation into lower-order modulation. We first introduce the confusion transmitter of Alice. Then, the deconfusion receiver of Bob is designed via dynamic programming (DP).
\subsubsection{Confusion Transmitter}
The principle of STD is to use identical symbols with varying durations from the confused modulation $\mathcal{B}$ to represent multiple symbols from the original modulation $\mathcal{A}$.
For the original modulation format $\mathcal{A}$, with an order $M$ where $M>N$, the alphabet is first divided into $N-1$ ordered subsets, denoted as
\begin{equation}
   {{\mathcal{A}}_n} = \big\{ a_1^{(n)},a_2^{(n)}, \ldots ,a_{{K_n}}^{(n)}\big\},\quad n \in {{\mathcal{I}}_{N - 1}}
\end{equation}
satisfying 
\begin{equation}
   \bigcap\limits_{n \in {{\mathcal{I}}_{N - 1}}} {{{\mathcal{A}}_n}}  = \O,~\bigcup\limits_{n \in {{\mathcal{I}}_{N - 1}}} {{{\mathcal{A}}_n}}  = {\mathcal{A}}
\end{equation}
where $K_n,~n\in {{\mathcal{I}}_{N - 1}}$ denotes the numbers of elements in $\mathcal{A}_n,~n\in {{\mathcal{I}}_{N - 1}}$.
Next, the symbols $a_i^{(n)},~i \in {{\mathcal{I}}_{{K_n}}},~n \in {{\mathcal{I}}_{N - 1}}$ are mapped to $b_n,~n \in {{\mathcal{I}}_{N - 1}}$ and last for $i$ symbol periods.
This means that different high-order modulation symbols are mapped to identical low-order modulation symbols, each of different length, that is,
\begin{equation}
\label{mapfunc1}
{x_l} = \underbrace {{b_n} \cdots {b_n}}_i,~{s_l} = a_i^{(n)},~i \in {{\mathcal{I}}_{{K_n}}},~n \in {{\mathcal{I}}_{N - 1}},~l \in {{\mathcal{I}}_{L}}
\end{equation}
where each $b_n$ lasts for one symbol period.
If two adjacent symbols in $\boldsymbol{s}$ belong to the same ${\mathcal{A}}_n$, this results in a longer sequence of identical symbols. 
It is challenging for Bob to accurately determine the actual duration of each symbol. 
For instance, with the symbol sequence $\underline{b_1b_1b_1}$, Bob cannot discern whether it represents $\underline{b_1}~\underline{b_1b_1}$, $\underline{b_1b_1}~\underline{b_1}$, or $\underline{b_1}~\underline{b_1}~\underline{b_1}$.
Therefore, if ${s_l},~l \in {{\mathcal{I}}_L} \setminus \{ 1\}$ belongs to the same subset as ${s_{l-1}}$ and ${s_{l-1}}$ is not mapped to $b_N$, $s_l$ is mapped to $b_N$ while retaining its original duration, that is
\begin{equation}
\label{mapfunc2}
{x_l} = \underbrace {{b_N} \cdots {b_N}}_i,~{s_l} = a_i^{(n)},~i \in {{\mathcal{I}}_{K_n}},~n \in {{\mathcal{I}}_{N - 1}},~l \in {{\mathcal{I}}_L}\setminus \{ 1\}.
\end{equation}
If they are not in the same subset, or if ${s_{l-1}}$ has already been mapped to $b_N$, the symbol is mapped normally using (\ref{mapfunc1}).
According to this rule, the original secure signal ${\boldsymbol{s}} = {\big[{s_1},{s_2}, \ldots ,{s_L}\big]^\mathsf{T}}$ is mapped to ${\boldsymbol{x}} = {[{{x}_1},{{x}_2}, \ldots ,{{x}_{L}}]^\mathsf{T}}$, where adjacent symbols in $\boldsymbol{x}$ are always distinct, and each symbol has a duration that is an integer multiple of the symbol period.
If analyzed using the original symbol period, the signal sequence actually observed is denoted as 
${\boldsymbol{x'}} = {[{{x'_1}},{{x'_2}}, \ldots ,{{x'_{J}}}]^\mathsf{T}}$ where $J$ is the number of symbols sampled after confusion.
The additional information contained in the higher-order modulation symbols, relative to the lower-order ones, is conveyed through the different durations of the symbols, resulting in $L < J$.
The high-to-low-order confusion transmitter is summarized in Algorithm 1.

\begin{algorithm}[t]
\label{alg01}
\caption{High-to-Low-Order Confusion Transmitter}
{\bf Input:} $\boldsymbol{s}$;\\
{\bf Initialize:} $\boldsymbol{x}=\boldsymbol{0}$;
\begin{algorithmic}[1] 
\State Update $x_1$ using (\ref{mapfunc1});
\State {\bf for} $l=2:L$ {\bf do}
    \State \quad {\bf if} $x_{l-1}$ is not mapped to $b_N$ and $\exists {{\mathcal{A}}_n},~n \in {{\mathcal{I}}_{N - 1}}$\\\quad such that ${s_{l - 1}},~{s_l} \in {{\mathcal{A}}_n}$ {\bf then}
    \State \qquad Update $x_l$ using (\ref{mapfunc2});
    \State \quad {\bf else}
    \State \qquad Update $x_l$ using (\ref{mapfunc1});
    \State \quad {\bf end if}
\State {\bf end for}   
\end{algorithmic}
{\bf Output:} $\boldsymbol{x}$.
\end{algorithm}

Considering communication efficiency and signal processing convenience, the order $M$ of the original modulation format $\mathcal{A}$ is typically a power of $2$. For the confused modulation format $\mathcal{B}$, its order $ N $ can also be set as a power of $2$, such as using common QAM or PSK modulations. However, this configuration prevents $M$ from being evenly divided into $N-1$ subsets, leading to significant variations in the probabilities of symbols appearing in the confused sequence $\boldsymbol{x'}$, which could alert Eve. Therefore, one might consider using flexible-order modulation formats like golden angle modulation (GAM) \cite{larsson2017golden}, setting $ N-1 $ as a power of $2$ to allow the symbols to appear with equal probability in $\boldsymbol{x'}$.
Specifically, the $n$-th constellation point of GAM has the complex amplitude
\begin{equation}
    {x_n} = {r_n}{e^{\jmath2\pi \varphi n}},\quad n \in {\mathcal{I}_N}
\end{equation}
where $\varphi  = 1 - \frac{{\sqrt 5  - 1}}{2}$, ${r_n} = {c_{{\text{disc}}}}\sqrt n ,n \in {\mathcal{I}_N}$ and ${c_{{\text{disc}}}} \triangleq \sqrt {\frac{{2\bar P}}{{N + 1}}} $.
Considering the power normalization constraint, $\bar P$ is typically set to $1$.

\subsubsection{Deconfusion Receiver}
Since Bob cannot know the duration of each symbol of $\boldsymbol{x}$ in advance, it samples using a unit symbol period, resulting in ${\boldsymbol{y'}} = {\big[y'_1,y'_2, \ldots ,y'_{J}\big]^\mathsf{T}}$, and the $j$-th symbol is given by
\begin{equation}
   y'_{j} = hx_{j} + \mu,\quad {j} \in {{\mathcal{I}}_{J}}.
\end{equation}
In $\boldsymbol{x}$ produced by Algorithm $1$, adjacent symbols are always different. Therefore, in 
$\boldsymbol{y'}$, we can restore the original symbol sequence by detecting points of symbol change and making a comprehensive judgment based on the time intervals of each change.
Given that the duration of each symbol is random, under the effects of fading channels and noise, misjudging symbols early in the sequence $\boldsymbol{y'}$ can lead to significant error propagation. For instance, misinterpreting $\underline{b_1}~\underline{b_2}$ as $\underline{b_1}~\underline{b_1}$ could cause two original symbols to be mistaken for one, resulting in symbol loss.
Thus, we design a deconfusion receiver of Bob based on the DP method to prevent error propagation.

First, Bob constructs a DP table ${{\boldsymbol{D}}_\text{T}}$ and a path table ${{\boldsymbol{P}}_\text{T}}$, where ${d_\text{T}}(j,l)$ and ${p_\text{T}}(j,l),~l \in {{\mathcal{I}}_L},~j \in {{\mathcal{I}}_{J}}$ represent the element at the $j$-th row and the $l$-th column of ${{\boldsymbol{D}}_\text{T}}$ and  ${{\boldsymbol{P}}_\text{T}}$, respectively.
For ${d_\text{T}}(j,l)$, it denotes the minimum dissimilarity cost for partitioning the first $j$ symbols into $l$ groups, and ${d_\text{T}}(0,0)$ is set to zero because no cost is associated with partitioning zero symbols into zero groups.
All the other elements in ${{\boldsymbol{D}}_\text{T}}$ are initialized to a very large value, representing an invalid or uncomputed state.
All the elements in ${{\boldsymbol{P}}_\text{T}}$ are set to zero.
Here, Bob is aware that a received frame of $\boldsymbol{y'}$ contains $L$ original symbols, and it also knows the maximum value of $K_n,~n\in {{\mathcal{I}}_{N-1}}$, i.e., ${K_{\max }}$, as previously agreed upon by Alice and Bob.
Then, Bob traverses the received symbol sequence $\boldsymbol{y'}$. If ${y'_{j}}$ forms a new group 
$l$, the total cost of the grouping is given by
\begin{equation}
\label{DPcost1}
   c_{j,1} = {d_\text{T}}(j - 1,l - 1) + c_1,~j \in {{\mathcal{I}}_{J}}
\end{equation}
where $c_1$ is a constant cost representing the cost of a single symbol group.
Furthermore, if ${y'_{j}}$ and the preceding $k-1$ symbols, totaling $k$ symbols, are grouped together, the total cost is
\begin{align}
\label{DPcost2}
   c_{j,k} &= {d_\text{T}}(j - k,l - 1) + {c'_{j,k}}\notag\\
   &j \in {{\mathcal{I}}_{J}}\setminus \{1\},\quad k\in {{\mathcal{I}}_{K_{\max}}}\setminus \{1\}
\end{align}
where 
\begin{align}
\hspace{5pt}{c'_{j,k}} &= \frac{1}{k}\sum\limits_{i = 0}^{k - 1} {\big| {{y'_{j - i}} - {{\tilde b_{j,k}}}} \big|} \notag \\
&j \in {{\mathcal{I}}_{J}}\setminus \{1\},\quad k\in {{\mathcal{I}}_{K_{\max}}}\setminus \{1\}
\end{align}
\begin{align}
\label{estsymbol}
\hspace{2.5em}&{{\tilde b_{j,k}}} = \mathop {\arg \min }\limits_{{b_n},~n \in {{\mathcal{I}}_N}} \bigg| {\frac{1}{k}\sum\limits_{i = 0}^{k - 1} {{y'_{j - i}}}  - {b_n}} \bigg|\notag\\
&\hspace{20pt}j \in {{\mathcal{I}}_{J}}\setminus \{1\},~k\in {{\mathcal{I}}_{K_{\max}}}\setminus \{1\}.
\end{align}
Note that when using (\ref{DPcost2}), it is required that $j \geqslant k$.
The DP table is updated by
\begin{equation}
\label{DPcost3}
{d_\text{T}}(j,l) = \min \left\{ {{c'_{j,k}}} \right\},~j \in {{\mathcal{I}}_{J}},~l \in {{\mathcal{I}}_L},~ k \in {{\mathcal{I}}_{{K_{\max }}}}.
\end{equation}
The path information is recorded in ${{\boldsymbol{P}}_\text{T}}$ to facilitate the reconstruction of the grouping sequence, that is
\begin{equation}
\label{DPpath}
{p_{\text{T}}}(j,l) = \mathop {\arg \min }\limits_{k \in {{\mathcal{I}}_{{K_{\max }}}}} \left\{ {{c'_{j,k}}} \right\},\quad j \in {{\mathcal{I}}_{J}},\quad l \in {{\mathcal{I}}_L}.
\end{equation}

Next, starting from ${p_\text{T}}(J,L)$, Bob backtracks using the stored path information to reconstruct the optimal grouping of symbols. The grouping path is determined by examining the transitions recorded in $\boldsymbol{P}_\text{T}$, which identify the number of symbols comprising a group.
Initially, Bob sets up ${{{\boldsymbol{\tilde x}}}_{{\text{inv}}}} = {\left[ {{{\tilde x}_{{\text{inv}},1}},{{\tilde x}_{{\text{inv}},2}}, \ldots ,{{\tilde x}_{{\text{inv}},L}}} \right]^\mathsf{T}}$ as the inverse sequence of ${\boldsymbol{\tilde x}} = {\left[ {{{\tilde x}_1},{{\tilde x}_2}, \ldots ,{{\tilde x}_L}} \right]^\mathsf{T}}$ post-deconfusion and sets $j=J,~l=L$.
Subsequently, Bob retrieves the element $k^*$ from $\boldsymbol{P}_\text{T}$, i.e., 
\begin{equation}
\label{computek}
k^* = {p_\text{T}}(j,l)
\end{equation}
and ${{\tilde x}_{{\text{inv}},l}}$ is given by
\begin{equation}
\label{xinverse}
{{\tilde x}_{{\text{inv}},l}} = \underbrace {{{\tilde b}_{j,{k^ * }}} \cdots {{\tilde b}_{j,{k^ * }}}}_{{k^ * }},\quad l \in {{\mathcal{I}}_L}
\end{equation}
where ${{{\tilde b}_{j,k^*}}}$ is computed using (\ref{estsymbol}).
Then, Bob updates $j$ as $j-k^*$ and $l$ as $l - 1$, iteratively computing to obtain the complete sequence ${{{\boldsymbol{\tilde x}}}_{{\text{inv}}}}$, and ${{\boldsymbol{\tilde x}}}$ is obtained by reversing ${{{\boldsymbol{\tilde x}}}_{{\text{inv}}}}$, i.e.,
\begin{equation}
\label{xtilde}
{{\tilde x}_l} = {{\tilde x}_{{\text{inv}},L - l + 1}},\quad l \in {{\mathcal{I}}_L}.
\end{equation}
The high-to-low-order deconfusion receiver is summarized in Algorithm $2$.

\begin{algorithm}[t]
\label{alg02}
\caption{High-to-Low-Order Deconfusion Receiver}
{\bf Input:} $\boldsymbol{y'}$,~$L$;\\
{\bf Initialize:} ${d_\text{T}}(0,0)=0$,~${d_\text{T}}(j,l)=10^6,~j \in {{\mathcal{I}}_{J}},~l \in {{\mathcal{I}}_L}$;\\
{\bf Initialize:} ${p_\text{T}}(j,l)=0,~j \in {{\mathcal{I}}_{J}},~l \in {{\mathcal{I}}_L}$;
\begin{algorithmic}[1] 
\State {\bf for} $j=1:J$ {\bf do}
    \State \quad {\bf for} $l=1:\min \{L,j\}$ {\bf do}
        \State \qquad Compute ${c'_{j,1}}$ using (\ref{DPcost1});
        \State \qquad {\bf for} $k=1:K_{\max}$ {\bf do}
            \State \qquad \quad {\bf if} $j \geqslant k$ {\bf then}
                \State \qquad \qquad Compute ${c'_{j,k}}$ using (\ref{DPcost2});
            \State \qquad \quad {\bf end if}   
        \State \qquad {\bf end for}  
        \State \qquad Update ${d_\text{T}}(j,l)$ using (\ref{DPcost3});
        \State \qquad Update ${p_\text{T}}(j,l)$ using (\ref{DPpath});
    \State \quad {\bf end for}  
\State {\bf end for}
\State $j=J$;
\State $l=L$;
\State {\bf while} $j > 0$ and $l > 0$ {\bf do}  
    \State \quad Compute $k^*$ using (\ref{computek});
    \State \quad Compute ${{\tilde x}_{{\text{inv}},l}}$ using (\ref{xinverse});
    \State \quad $j= j- k^*$;
    \State \quad $l = l - 1$;
\State {\bf end while}
\State Compute ${{\boldsymbol{\tilde x}}}$ using (\ref{xtilde});
\end{algorithmic}
{\bf Output:} $\boldsymbol{x}$.
\end{algorithm}

\emph{Remark 2:}
The proposed high-to-low-order confusion shares conceptual similarities with variable-length source coding, such as Huffman coding, where symbols are mapped to unequal-length representations. However, directly applying such coding to modulation confusion leads to substantial spectral efficiency loss, since prefix-free constraints force long sequences to dominate transmission time. Future work may explore source-coding-inspired designs under explicit rate constraints to better balance security and spectral efficiency.

\emph{Remark 3:}
In high-to-low order modulation mapping, fixed-length mapping simplifies receiver design but suffers from reduced spectral efficiency and security compared with variable-length mapping. This is because each high-order symbol is tied to a fixed low-order sequence, causing the statistical distribution of low-order symbols to directly reflect that of the original modulation. Such correlations can be readily exploited by modulation classification algorithms, and even with uniform source distributions, fixed mappings introduce sequence regularities that weaken resistance to classification.

\subsection{Analysis of the Single-Antenna Schemes}
In this section, we first analyze the security of the proposed scheme.
Then, we provide a detailed analysis and comparison of the advantages and disadvantages of two confusion schemes.

For the low-to-high-order confusion scheme, a significant advantage is the ability to balance the communication performance and the security by adjusting the mapping probabilities. Compared to the rigid switching in adaptive modulation techniques, this scheme offers greater flexibility in adapting to channel conditions. However, a drawback is the inevitable decrease in BER performance due to the increase in the modulation order.

For the high-to-low-order confusion scheme, the advantage is the improvement in BER performance. One reason is that a reduction in order leads to a sparser constellation diagram, enhancing noise resistance. Another significant reason is the error detection capability of the proposed receiver based on DP method. As described in Section II.B, the same symbol in $\boldsymbol{y'}$ appears consecutively a maximum of $K_{\max}$ times. If it appears more frequently, the Algorithm $2$ will further group them according to the optimal path and perform a combined decision demodulation for multiple symbols in the same group, as demonstrated in equation (\ref{estsymbol}).
However, a drawback of the high-to-low-order confusion scheme is the loss of spectral efficiency. For example, when disguising a $8$-order modulation to a $4$-order modulation, the spectral efficiency decreases from $3\text{bit}/\text{s}/\text{Hz}$ to $2\text{bit}/\text{s}/\text{Hz}$.

The security of the proposed MOC schemes mainly stem from misleading the modulation classifier of Eve through SRM and STD. By optimizing the symbol mapping probabilities or varying the symbol durations, the transmitted signal distribution becomes indistinguishable from that of another modulation order, while the legitimate receiver, equipped with the mapping or duration information, can reliably recover the original symbols. 
However, the security of the proposed schemes may be reduced if Eve obtains partial prior knowledge of the mapping or duration distribution. Moreover, there is an inherent trade-off between security and communication performance, as excessive confusion may result in reduced spectral efficiency or increased bit error rate. These limitations suggest that future work could enhance the robustness of this scheme by introducing dynamic mapping, time-frequency hybrid confusion, and multi-antenna cooperation.

In the single-antenna systems, the proposed two MOC schemes require Bob to design a specific receiver and agree on some parameter settings with Alice in advance, which may not be permissible in some scenarios, such as when the computational resources of Bob are limited. To address this issue, the next section introduces multi-antenna MOC schemes that enables secure communication even when Bob lacks prior information.

\section{Modulation Order Confusion for Multi-Antenna Communication}
In this section, we first introduce a multi-antenna communication system. Then, a MOC scheme based on signal fusion is introduced.

\subsection{Multi-Antenna System Model}

Consider a point-to-point communication system, which consists of a transmitter (Alice) with ${T_{\text{A}}}$ antennas, a single-antenna legitimate receiver (Bob), and a passive eavesdropper (Eve) with $T_{\text{E}}$ antennas. 
The original secure symbol is $s$, ${s} \in {\mathcal{A}}$, the transmit signal after confusion is
${{\boldsymbol{x}}} = {\big[{x_{1}},{x_{2}}, \ldots ,{x_{{T_{\text{A}}}}}\big]^\mathsf{T}}$ where ${x_{{t_{\text{A}}}}},~{t_{\text{A}}} \in {{\mathcal{I}}_{{T_{\text{A}}}}}$ represents the symbol transmitted by the ${{t_{\text{A}}}}$-th antenna of Alice, and ${x_{{t_{\text{A}}}}} \in {{\mathcal{B}_{{t_{\text{A}}}}}}$. 
Here, ${{\mathcal{B}_{{t_{\text{A}}}}}},~{t_{\text{A}}} \in {{\mathcal{I}}_{{T_{\text{A}}}}}$ is the alphabet of the ${{t_{\text{A}}}}$-th antenna, and ${N_{{t_{\text{A}}}}},~{t_{\text{A}}} \in {{\mathcal{I}}_{{T_{\text{A}}}}}$ denotes the modulation order of ${{\mathcal{B}_{{t_{\text{A}}}}}}$.
The received symbol at Bob is
\begin{align}
{y} = {\boldsymbol{hW}}{{\boldsymbol{x}}}  + \mu
\end{align} 
where ${\boldsymbol{W}} = {\text{diag}}\left\{ {\boldsymbol{w}} \right\} \in {\mathbb{C}^{T_\text{A} \times T_\text{A}}}$ denotes the beamformer matrix; 
${\boldsymbol{w}} = \left[ {{w_1},{w_2}, \ldots ,{w_{T_\text{A}}}} \right] \in {\mathbb{C}^{T_\text{A}}}$ is the beamformer vector; 
${\boldsymbol{h}} = \left[ {{h_1},{h_2}, \ldots ,{h_{T_\text{A}}}} \right] \in {\mathbb{C}^{T_\text{A}}}$ represents the channels form Alice to Bob;
$\mu$ is the noise at Bob following ${\mathcal{C}\mathcal{N}}(0,{\sigma  ^2})$.
The received signal of Bob is
\begin{equation}
{{\boldsymbol{z}}} = {\boldsymbol{GW}}{{\boldsymbol{x}}} + \boldsymbol{\nu}
\end{equation}
where ${\boldsymbol{G}} \in {\mathbb{C}^{T_{\text{E}} \times T_{\text{A}}}}$ is the channel responses from Alice to Eve;
${\boldsymbol{\nu }}$ represent the noise at Eve, i.e.,  $\boldsymbol{\nu}  \sim \mathcal{CN}(0,{\gamma ^2})$.
The channel $\boldsymbol{G}$ remains entirely obscure to the legitimate nodes since Eve does not provide any information. 
 
\subsection{Beamformer Design}
In this section, we propose the beamformer design strategy.
Our goal is for the signals from each antenna of Alice to directly aggregate at Bob without channel interference, thereby facilitating the design of the confusion transmitter.
Thus, we set the constraint
\begin{equation}
 {\boldsymbol{h}}{\boldsymbol{W}} = {{\boldsymbol{I}}_{T_{\text{A}}}}
\end{equation}
where ${{\boldsymbol{I}}_{T_{\text{A}}}} = \underbrace {\left[ {1,1, \ldots ,1} \right]}_{T_{\text{A}}}$,
we have
\begin{align}
y &= {\boldsymbol{hW}}{{\boldsymbol{x}}}  + \mu  \\
\label{multi_Bob1}
&=  \sum\limits_{{t_{\text{A}}} = 1}^{T_{\text{A}}} {{x_{t_{\text{A}}}}}  + \mu.
\end{align}
Here, we assume that $ {\boldsymbol{h}}$ is available to Alice, which is a common assumption in the physical layer security literature. As a result, the beamformer vector $\boldsymbol{w}$  can be easily designed using 
\begin{align}
{w_{{t_{\text{A}}}}} = \frac{1}{{{h_{{t_{\text{A}}}}}}},\quad {t_{\text{A}}} \in {\mathcal{I}_{{T_{\text{A}}}}}.
\end{align}

\begin{figure*}[t]
\centering
\includegraphics[width=12.2cm]{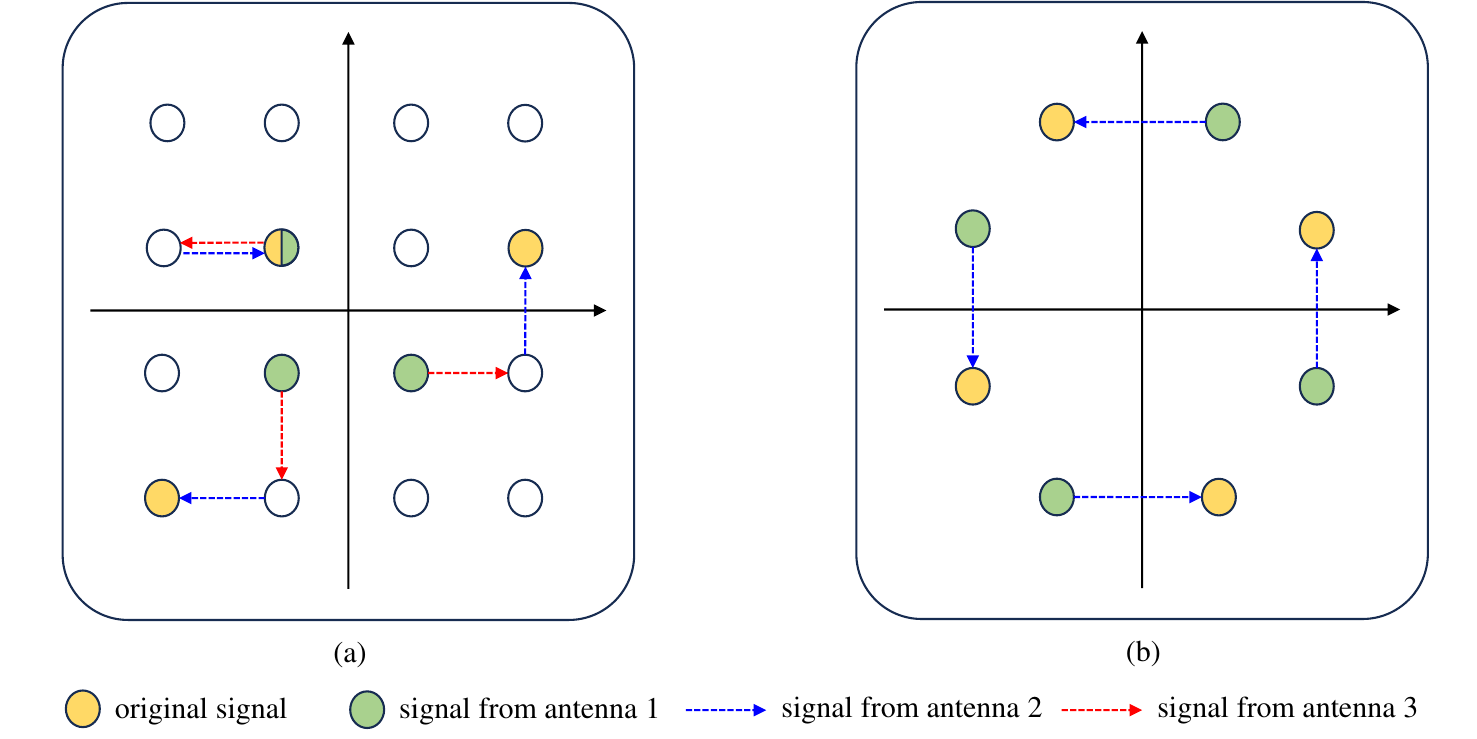}
\caption{Illustration of the symbol design strategy based on CPD.
In subfigure (a), the antenna $1$ transmits signals from the set $\mathcal{B}_1 = \{ 1+1\jmath, 1-1\jmath, -1+1\jmath, -1-1\jmath\} $, while antennas $2$ and $3$ transmit signals from identical sets $ \mathcal{B}_2 = \mathcal{B}_3 = \{2,\,-2,\,2\jmath,\,-2\jmath\} $.
In subfigure (b), the antenna $1$ transmits symbols from the set \( {\mathcal{B}_1} = \{ {e^{\frac{3}{8}\pi }},{e^{\frac{7}{8}\pi }},{e^{ - \frac{1}{8}\pi }},{e^{ - \frac{5}{8}\pi }}\}  \), while antenna $2$ uses the set \({\mathcal{B}_2} = \{ \sqrt {2 - \sqrt 2 } ,-\sqrt {2 - \sqrt 2 } ,\sqrt {2 - \sqrt 2 }\jmath ,-\sqrt {2 - \sqrt 2 }\jmath \}  \).}
\label{constellation}
\end{figure*}

\subsection{Low-to-High-Order Confusion}
In this section, we propose the confusion transmitter based on the Taylor expansion.
At Alice, the original secure signal \( s \) undergoes a nonlinear transform to produce \( f(s) \). One antenna transmits \( f(s) \), while the remaining antennas transmit the negatives of the nonlinear terms from the Taylor series expansion of \( f(s) \). When all the antenna signals are aggregated at Bob, they result in a linear term of the original secure signal \( s \).
Specifically, for a nonlinear function \( f(s) \) that is differentiable at $s_0$,
the first antenna transmits the nonlinear signal $f(s)$ directly, i.e.,
\begin{equation}
\label{Antenna1}
{x_1} = f(s).
\end{equation}
The antenna ${{t_{\text{A}}}},~{t_{\text{A}}} \in {\mathcal{I}_{{T_{\text{A}}}}} \setminus \{ 1\} $
transmits
\begin{align}
\label{AntennatA}
&{x_{{t_{\text{A}}}}} =  - \sum\limits_{n \in {\mathcal{M}_{{t_{\text{A}}}}}} {\frac{{{\lambda _n}}}{{n!}}{{(s - {s_0})}^n}} ,~~{t_{\text{A}}} \in {\mathcal{I}_{{T_{\text{A}}}}} \setminus \{ 1\}  \notag \\
&\bigcap\limits_{{t_{\text{A}}} \in {\mathcal{I}_{{T_{\text{A}}}}} \setminus \{ 1\} } {{\mathcal{M}_{{t_{\text{A}}}}}}  = \O,~ \bigcup\limits_{{t_{\text{A}}} \in {\mathcal{I}_{{T_{\text{A}}}}} \setminus \{ 1\} } {{\mathcal{M}_{{t_{\text{A}}}}}}  = {\mathcal{I}_\infty } \setminus \{ 1\} 
\end{align}
where 
\begin{equation}
{\lambda _n} = \mathop {\lim }\limits_{s \to {s_0}} \frac{{{{\text{d}}^n}f(s)}}{{{\text{d}}{s^n}}},\quad n \in {\mathcal{I}_\infty }.
\end{equation}
Thus, the received signal of Bob is given by
\begin{align}
\label{multi_Bob3}
y &= \sum\limits_{{t_{\text{A}}} = 1}^{{T_{\text{A}}}} {{x_{{t_{\text{A}}}}}}  + \mu \\
\label{multi_Bob4}
&= f(s) - \sum\limits_{{t_{\text{A}}} = 1}^{{T_{\text{A}}}} {\sum\limits_{n \in {\mathcal{M}_{{t_{\text{A}}}}}}\hspace{-5pt} {\frac{{{\lambda _n}}}{{n!}}{{(s - {s_0})}^n}} }  + \mu   \\
\label{multi_Bob5}
\begin{split}
&=f({s_0}) +\hspace{-5pt} \sum\limits_{n \in {\mathcal{I}_\infty }} {\frac{{{\lambda _n}}}{{n!}}{{(s - {s_0})}^n}} - \hspace{-1em}\sum\limits_{n \in {\mathcal{I}_\infty } \setminus \{ 1\} } \hspace{-3pt}{\frac{{{\lambda _n}}}{{n!}}{{(s - {s_0})}^n}}  + \mu   
\end{split}\\
&= {\lambda _1}s + f({s_0}) - {\lambda _1}{s_0} + \mu 
\end{align}
where (\ref{multi_Bob3}) follows from (\ref{multi_Bob1});
(\ref{multi_Bob4}) follows from (\ref{Antenna1}) and (\ref{AntennatA});
(\ref{multi_Bob5}) follows Taylor's Theorem.
Then, Bob can recover the original secure signal $s$ by filtering out the direct current (DC) and normalizing the power.
Note that the transmit signal at each antenna typically contains a DC component, which is essential for Bob to recover $s$. Although the antennas need to remove the DC component before transmitting, Bob can still recover $s$ through simple methods such as DC biasing. 
For brief illustration, we omit this factor in the formula derivation. 
Typically, the mean of $s$ is zero, so we can set $s_0=0$, meaning that $f(s)$ is
expanded around $s=0$ and Bob can reduce the computational overhead. 
In fact, different values of $s_0$ can bring additional degrees of secrecy freedom.

In reality, Alice cannot transmit an infinite number of Taylor series terms with a finite number of antennas. Typically, transmitting only the largest several terms is sufficient for Bob to receive the original secure signal \( s \) almost perfectly. This will be discussed in the simulation section later.  
The antennas can randomly allocate signal terms to allow the original $M$-order modulation to be disguised as $N$-order modulation on each antenna.
When the sets \( {\mathcal{M}_{{t_{\text{A}}}}},~{t_{\text{A}}} \in {\mathcal{I}_{{T_{\text{A}}}}} \setminus \{ 1\}  \) are fixed, \( N = M{T_{\text{A}}}\).

Furthermore, the function \( f(s) \) requires analysis of its convergence radius, which is defined as the radius of the largest disc centered at the series convergence center within which the series converges. 
For \( f(s) = \arctan s \), as an example, the radius is determined to be $1$. As a result, for PSK modulation, normalizing the power results in all constellation points residing on the unit circle, ensuring the convergence of the Taylor series expansion. In contrast, with QAM modulation, several symbols extend beyond the unit circle even after power normalization. This necessitates amplitude normalization to guarantee that all constellation points remain inside or on the unit circle.
For a more detailed discussion on the series-expansion-based secure communication scheme, please refer to our previous work\cite{wang2025secure}.

\subsection{High-to-Low-Order Confusion}

In this section, we propose a confusion transmitter based on constellation path design (CPD) that disguises the signals transmitted from each antenna of Alice  as lower-order modulations.
Specifically, when transmitting an original secure signal \( s \) from a modulation set \( \mathcal{A} \) of order \( M \), the first antenna of Alice transmits a signal \( {x_1} \in {\mathcal{B}_1} \) with ${N_1} < M$. The signals from the remaining antennas adjust the position of \( x_1 \) in the constellation diagram to eventually shift it to the location of \( s \). 
The signal transmitted by the antenna ${t_{\text{A}}},~{t_{\text{A}}} \in {\mathcal{I}_{{T_{\text{A}}}}} \setminus \{ 1\} $, denoted as \( {x_{{t_{\text{A}}}}},~{t_{\text{A}}} \in {\mathcal{I}_{{T_{\text{A}}}}} \setminus \{ 1\} \), enables \( x_1 \) to undergo a well-designed shift. 
Note that ${N_{{t_{\text{A}}}}} < M,~{t_{\text{A}}} \in {\mathcal{I}_{{T_{\text{A}}}}} \setminus \{ 1\} $. These shifts form a set \( {\mathcal{B}_{{t_{\text{A}}}}},~{t_{\text{A}}} \in {\mathcal{I}_{{T_{\text{A}}}}} \setminus \{ 1\}  \), serving as a set of modulations intended to confuse Eve. 

As an illustrative example, Figure $\ref{constellation}$ demonstrates how 16QAM and 8PSK modulations can be disguised as lower-order modulations through CPD.
In Figure $\ref{constellation}$(a), each constellation point in the 16QAM modulation can be obtained by starting from a symbol in \( \mathcal{B}_1 \) and applying offsets using symbols from \( \mathcal{B}_2 \) and \( \mathcal{B}_3 \). 
Even if Eve uses blind source separation to obtain signals transmitted from the three antennas, it can only determine that each antenna employs a certain type of 4-order digital modulation, and thus cannot decode the original secure signal. 
Similarly, in Figure $\ref{constellation}$(b), the antenna $1$ transmits symbols from the set \( {\mathcal{B}_1}  \), while antenna $2$ uses the set \({\mathcal{B}_2}  \).
The two antennas collaborate to transmit 8PSK modulation symbols based on their respective 4-order symbol sets. 

\emph{Remark 4:}
In the multi-antenna system, we design the transmit symbols for each antenna of Alice such that the order of each transmit symbol set is higher or lower than the original secure symbol set, yet these symbols can be aggregated at Bob to recover the original secure symbols. 
Consequently, even if Eve captures signals from some antennas of Alice  using advanced blind source separation techniques, it still cannot accurately identify the modulation format. Moreover, due to the amplitude and phase ambiguities inherent in blind source separation, it cannot simply sum the signals from various antennas to recover the original secure signal.

\section{Application of the MOC approach for RIS-Assisted Multi-Antenna Communication}
In recent years, multi-antenna systems have evolved into various new communication paradigms, among which reconfigurable intelligent surface (RIS) have attracted significant research attention.
In this section, we first introduce a RIS-assisted multi-antenna communication system. Then, we provide the design of the RIS and beamformer.

\subsection{RIS-Assisted Multi-Antenna System Model}

\begin{figure}[t]
\centering
\includegraphics[width=8.5cm]{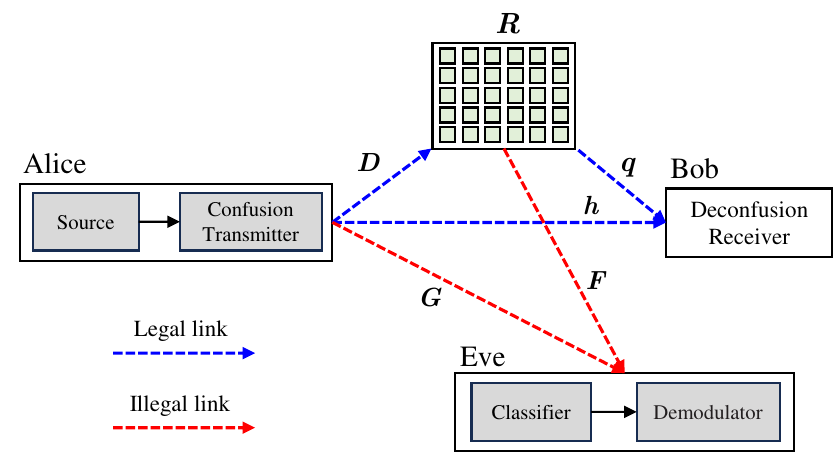}
\caption{Illustration of the RIS-assisted multi-antenna system.}
\label{systemmodel2}
\end{figure}

Consider a RIS-aided communication system, which consists of a transmitter (Alice) with ${T_{\text{A}}}$ antennas, a single-antenna legitimate receiver (Bob) , a passive eavesdropper (Eve) with $T_{\text{E}}$ antennas and a RIS with $V$ reflecting elements, as shown in Figure \ref{systemmodel2}. 
The received symbol at Bob when there is and isn't a direct link between Alice and Bob are $y_{1}$ and $y_{2}$ respectively, given by
\begin{align}
\label{multiBob}
{y_{1}} &= {\boldsymbol{hW}}{{\boldsymbol{x}}} + {\boldsymbol{qRDW}}{{\boldsymbol{x}}} + \mu_{1}\\
\label{multiBob2}
{y_{2}} &= {\boldsymbol{qRDW}}{{\boldsymbol{x}}} + \mu_{2}
\end{align} 
where the matrices ${\boldsymbol{D}} \in {\mathbb{C}^{V \times T_\text{A}}}$ is the CSI from Alice to the RIS;
${\boldsymbol{R}} = {\text{diag}}\left\{ {\boldsymbol{r}} \right\} \in {\mathbb{C}^{V \times V}}$ is the phase shift matrix of the RIS, and we have ${\boldsymbol{r}} = \left[ {{r_1},{r_2}, \ldots ,{r_V}} \right] \in {\mathbb{C}^V}$ where ${r_v} = {{\rho _v}e^{\jmath{\theta _v}}},~v \in {{\mathcal{I}}_V}$,
${\rho _v}$ and ${\theta _v}$ are the amplitude and the phase shift of the $v$-th reflecting element of the RIS.\footnote{To study a generalized model, we consider an ideal RIS case, i.e., the amplitude and the phase of each RIS element can be independently and continuously controlled \cite{zheng2018ultra}. Furthermore, we consider a passive RIS, i.e., ${\rho _v} \leqslant 1,~v \in {{\mathcal{I}}_V}$.}
The vector ${\boldsymbol{q}} = \left[ {{q_1},{q_2}, \ldots ,{q_V}} \right] \in {\mathbb{C}^N}$ represents the channels from the RIS to Bob.
$\mu_{1}$ and $\mu_{2}$ are the noise at Bob following ${\mathcal{C}\mathcal{N}}(0,{\sigma  ^2})$.
The received signal of Eve is
\begin{equation}
{{\boldsymbol{z}}} = {\boldsymbol{GW}}{{\boldsymbol{x}}} + {\boldsymbol{FRDW}}{{\boldsymbol{x}}} + \boldsymbol{\nu}
\end{equation}
where ${\boldsymbol{F}} \in {\mathbb{C}^{T_{\text{E}} \times V}}$ is the channel responses from the RIS to Eve;
${\boldsymbol{\nu }}$ represent the noise at Eve, i.e.,  $\boldsymbol{\nu}  \sim \mathcal{CN}(0,{\gamma ^2})$.
We assume that $\boldsymbol{D}$, $\boldsymbol{q}$ and $\boldsymbol{h}$ are perfectly known to Alice and unknown to Bob.\footnote{The channels $\boldsymbol{D}$ and $\boldsymbol{q}$ can be separately estimated at Alice via a two stage channel estimation method \cite{he2019cascaded}.
Even though $\boldsymbol{D}$ and $\boldsymbol{q}$ might be estimated imprecisely as $\boldsymbol{D'}$ and $\boldsymbol{q'}$, where ${\boldsymbol{D'}} = {\boldsymbol{D}}\boldsymbol{\Phi} $ and ${\boldsymbol{q'}} = {\boldsymbol{\Phi} ^{ - 1}}{\boldsymbol{q}}$ with $\boldsymbol{\Phi}$ as a full-rank diagonal matrix, the suggested reflection designs remain effective because ${\boldsymbol{qRD}} = {\boldsymbol{q'RD'}}$. For additional information, please consult \cite{he2019cascaded}.}
Furthermore, it is assumed that Eve possesses complete understanding of $\boldsymbol{G}$ and performs maximum-likelihood detection using $\boldsymbol{G}$. We further assume that Eve is unaware of $\boldsymbol{F}$ for several reasons. Initially, the RIS does not emit any public signal detectable by Eve. Additionally, the reflection coefficients of RIS are subject to continuous variations, complicating the ability of Eve to monitor $\boldsymbol{F}$. Lastly, confidential channel estimation techniques \cite{yan2018secret} may be employed to safeguard the information pertaining to the legitimate channel.
Note that in the RIS-assisted multi-antenna system, the transmit signals at each antenna of Alice are the same as those described in Sections III.C and III.D. The difference lies in the need for additional specialized design of the beamformer and the RIS.

\subsection{Beamformer and RIS Design}
In this section, we propose a beamformer and RIS design strategy.
When there is a direct link between Alice and Bob, (\ref{multiBob}) can be written as
\begin{equation}
{y_{1}} = \left( {{\boldsymbol{h}} + {\boldsymbol{qRD}}} \right){\boldsymbol{Wx}} + \mu_1. 
\end{equation}
Then, in the the constraint
\begin{equation}
\label{constraint1}
\left( {{\boldsymbol{h}} + {\boldsymbol{qRD}}} \right){\boldsymbol{W}} = {{\boldsymbol{I}}_{T_{\text{A}}}}
\end{equation}
where ${{\boldsymbol{I}}_{T_{\text{A}}}} = \underbrace {\left[ {1,1, \ldots ,1} \right]}_{T_{\text{A}}}$,
we have
\begin{equation}
{y_{1}}  =  \sum\limits_{{t_{\text{A}}} = 1}^{T_{\text{A}}} {{x_{t_{\text{A}}}}}  + \mu_1.
\end{equation}
Consider the power budget ${\Gamma _{t_{\text{A}}}}$ of the ${t_{\text{A}}}$-th antenna, the RIS design problem is formulated as 
\begin{subequations}
  \label{Problem3}
  \begin{align} 
  \label{Problem3a}
{\mathcal{P}2}:~~\mathop {{\text{Find}}}\hspace{0.5em}&\boldsymbol{R}\\
    \label{Problem3b}{\text{s.t.}}\hspace{0.5em}&{\left| {{w_{{t_{\text{A}}}}}} \right|^2} \leqslant {\Gamma _{{t_{\text{A}}}}},\quad {t_{\text{A}}} \in {{\mathcal{I}}_{T_{\text{A}}}}\\
    \label{Problem3c}&{\rho _v} \leqslant 1,\quad v \in {{\mathcal{I}}_V}
  \end{align}
\end{subequations}
where
\begin{equation}
\label{beamformer1}
{w_{t_{\text{A}}}} = \frac{1}{{{h_{t_{\text{A}}}} + {\boldsymbol{qR}}{{\boldsymbol{D}}_{:,{t_{\text{A}}}}} }},\quad {t_{\text{A}}} \in {{\mathcal{I}}_{T_{\text{A}}}}
\end{equation}
and ${{{\boldsymbol{D}}_{:,{t_{\text{A}}}}}}$ is the ${t_{\text{A}}}$-th column of the matrix $\boldsymbol{D}$.
The equation (\ref{beamformer1}) is driven from (\ref{constraint1}).
The problem in $\mathcal{P}2$ is non-convex as the constraints contain the entangled term.
To solve the problem in $\mathcal{P}2$, we transform (\ref{Problem3b}) using the absolute value inequality and get
\begin{equation}
\label{beamformer2}
    \left| {{\boldsymbol{qR}}{{\boldsymbol{D}}_{:,{t_{\text{A}}}}}} \right| \geqslant \frac{1 }{{\sqrt {{\Gamma _{t_{\text{A}}}}} }} - \left| {{h_{t_{\text{A}}}}} \right|,\quad {t_{\text{A}}} \in {{\mathcal{I}}_{T_{\text{A}}}}.
\end{equation}
Then, we relax (\ref{beamformer2}) to a constraint on the real part and recast $\mathcal{P}2$ as 
\begin{subequations}
  \label{Problem4}
  \begin{align} 
  \label{Problem4a}\hspace{-7pt}
{\mathcal{P}3}:~~\mathop {{\text{Find}}}\hspace{0.5em}&\boldsymbol{R}\\
    \label{Problem4b}{\text{s.t.}}\hspace{0.5em}&\Re \left\{ {{\boldsymbol{qR}}{{\boldsymbol{D}}_{:,{t_{\text{A}}}}}} \right\} \geqslant \frac{1 }{{\sqrt {{\Gamma _{t_{\text{A}}}}} }} - \left| {{h_{t_{\text{A}}}}} \right|,~{t_{\text{A}}} \in {{\mathcal{I}}_{T_{\text{A}}}}\\
    \label{Problem4c}&{\rho _v} \leqslant 1,~v \in {{\mathcal{I}}_V}.
  \end{align}
\end{subequations}
The problem in $\mathcal{P}3$ is a convex problem that can be solved using CVX.
Note that although CVX can solve the feasibility problems where no objective function is specified and only constraints are defined, the solution process does not involve random sampling. Instead, the solver returns a deterministic feasible point chosen by the optimizer, typically located at an interior or boundary region, and not selected at random.

In the non-direct-link case, the constraint needs to be rephrased as 
\begin{equation}
\label{eq21}
{\boldsymbol{qRDW}} =  {{\boldsymbol{I}}_{T_{\text{A}}}}
\end{equation}
and the feasibility problem is formulated as 
\begin{subequations}
  \label{Problem5}
  \begin{align} 
  \label{Problem5a}\hspace{-7pt}
{\mathcal{P}4}:~~~~\mathop {{\text{Find}}}\hspace{0.5em}&\boldsymbol{R}\\
    \label{Problem5b}{\text{s.t.}}\hspace{0.5em}&\Re \left\{ {{\boldsymbol{qR}}{{\boldsymbol{D}}_{:,{t_{\text{A}}}}}} \right\} \geqslant \frac{1 }{{\sqrt {{\Gamma _{t_{\text{A}}}}} }} ,\quad {t_{\text{A}}} \in {{\mathcal{I}}_{T_{\text{A}}}}\\
    \label{Problem5c}&{\rho _v} \leqslant 1,\quad v \in {{\mathcal{I}}_V}
  \end{align}
\end{subequations}
which can also be solved by CVX.
Thus, we also have
\begin{equation}
{y_{2}}  =  \sum\limits_{{t_{\text{A}}} = 1}^{T_{\text{A}}} {{x_{t_{\text{A}}}}}  + \mu_2.
\end{equation}

\emph{Remark 5:}
Compared to the conventional multi-antenna system discussed in Section III, RIS introduces multiplicative randomness \cite{9luo2021reconfigurable}, which further prevents Eve from obtaining any useful information.

\section{Numerical Results}
In this section, we evaluate the proposed security scheme in terms of both the error performance and the security effectiveness. The key findings are as follows.

\begin{figure}[t]
\hspace{-20pt}
\includegraphics[width=10.1cm]{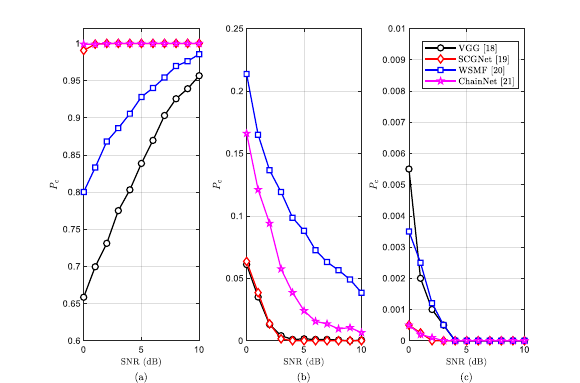}
\caption{The classification accuracy is evaluated w.r.t SNR for low-to-high-order confusion in the single-antenna system. In the simulation, QPSK is disguised as 16QAM. The three subplots correspond to different symbol mapping probabilities, i.e., (a) \( {\boldsymbol{p}} = [0, 0, 0, 1] \), (b) \( {\boldsymbol{p}} = [0.1, 0.2, 0.3, 0.4] \), and (c) \( {\boldsymbol{p}} = [0.25, 0.25, 0.25, 0.25] \). Additionally, we assume that Eve employs four types of deep-learning based modulation classifiers, i.e., VGG\cite{VGGo2018over}, SCGNet\cite{SCGNettunze2020sparsely}, WSMF\cite{WSMFqi2020automatic},  and ChainNet\cite{ChainNethuynh2020chain}.}
\label{Result1}
\end{figure}

\begin{figure}[t]
\hspace{-12pt}
\includegraphics[width=9.3cm]{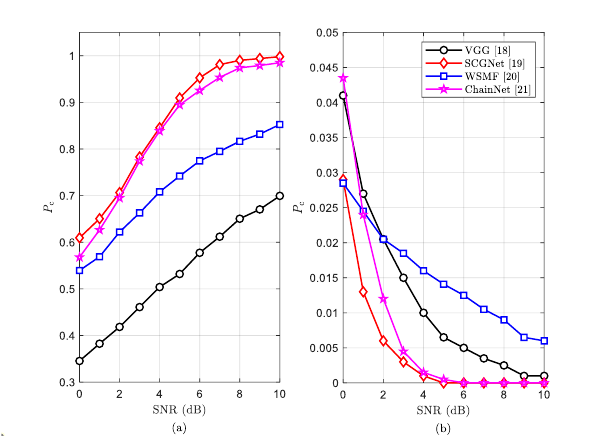}
\caption{The classification accuracy is evaluated w.r.t SNR for high-to-low-order confusion in the single-antenna system. In the simulation, 16PSK is disguised as 9GAM. 
The two subplots correspond to the cases before and after MOC, i.e., (a) without confusion, and (b) after confusion. 
Here, VGG\cite{VGGo2018over}, SCGNet\cite{SCGNettunze2020sparsely}, WSMF\cite{WSMFqi2020automatic},  and ChainNet\cite{ChainNethuynh2020chain} are also assumed to be employed at Eve.}
\label{Result2}
\end{figure}

\emph{\textbf{Observation 1:}
In single-antenna systems, both the proposed low-to-high and high-to-low modulation confusion schemes effectively prevent Eve from identifying the modulation format. Moreover, as the SNR increases, the proposed scheme demonstrates improved resistance against modulation classification
 (cf. Figures~$\ref{Result1}$ and $\ref{Result2}$).
} 

We evaluate the security of the proposed MOC scheme in the single-antenna system. The probability of the correct classification, $ P_\text{c} $, is used as the metric of the security performance. 
Since the proposed scheme requires a customized transmitter design, we do not use publicly available modulation datasets but instead generated our own. The dataset includes $10$ modulation schemes, i.e., 2FSK, 4FSK, MSK, GMSK, BPSK, QPSK, 8PSK, 16PSK, 16QAM, and 32QAM, with a roll-off factor of $0.35$ and an oversampling rate of $8$. To simulate channel effects, we apply random phase rotations to the samples after power normalization. 
In Figure~$\ref{Result1}$, QPSK is disguised as 16QAM through SRM. 
The resulting symbol set is divided into four rotationally symmetric subsets. 
Specifically, each symbol in the original QPSK is mapped to a random symbol within a designated quadrant of the 16QAM constellation. For example, the symbol $1+0\jmath$ is mapped to one of the mapping subset $\{1+1\jmath,~3+1\jmath,~1+3\jmath,~3+3\jmath\}$. 
In subplot (a), the mapping probability is \( {\boldsymbol{p}} = [0, 0, 0, 1] \), meaning the disguised signal retains the characteristics of a 4-order modulation, essentially without modulation confusion. As a result, all the four deep-learning based classifiers, VGG\cite{VGGo2018over}, SCGNet\cite{SCGNettunze2020sparsely}, WSMF\cite{WSMFqi2020automatic}, and ChainNet\cite{ChainNethuynh2020chain}, can accurately recognize the modulation format, with classification performance improving as SNR increases. 
In subplot (b), the mapping probability is \( {\boldsymbol{p}} = [0.1, 0.2, 0.3, 0.4] \), indicating that QPSK is partially mapped into a 16-order modulation, although the distribution of constellation points is nonuniform. Compared to subplot (a), the classification performance of the four classifiers noticeably declines, and further degrades with increasing SNR. This is because as the noise power decreases, the deviation of the disguised signal from the typical QPSK pattern becomes more apparent. In subplot (c), the mapping probability is \( {\boldsymbol{p}} = [0.25, 0.25, 0.25, 0.25] \), which means the disguised signal fully resembles a conventional 16QAM modulation. As a result, the classifier performance further deteriorates, with the classification accuracy dropping to zero as SNR increases. This indicates that under such conditions, Eve is unable to correctly identify the original modulation format. 
In Figure~$\ref{Result2}$, 16PSK is disguised as 9GAM using STD. 
In subplot (a), order confusion is not applied, so all the four classifiers exhibit normal modulation classification performance. In subplot (b), the classifiers show a significantly degraded classification accuracy for the disguised signal, and the security performance improves as the SNR increases. 

\begin{figure}[t]
\hspace{-10pt}
\includegraphics[width=9.7cm]{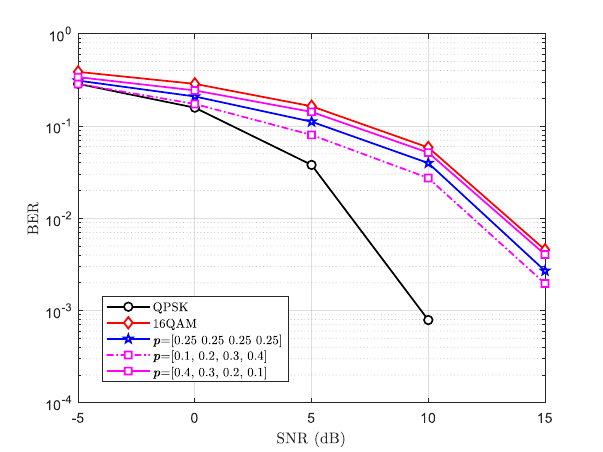}
\caption{The BER of Bob is evaluated w.r.t SNR for low-to-high-order confusion in the single-antenna system. In the simulation, QPSK is disguised as 16QAM. We simulate the BER under two different symbol mapping probabilities, \( {\boldsymbol{p}} = [0.1, 0.2, 0.3, 0.4] \) and \( {\boldsymbol{p}} = [0.25, 0.25, 0.25, 0.25] \). The BER of QPSK and 16QAM is also simulated as benchmarks for comparison.}
\label{Result3}
\end{figure}

\emph{\textbf{Observation 2:}
In the single-antenna system, under the proposed low-to-high MOC scheme, the BER at Bob lies between that of the original low-order modulation and the target high-order modulation. Moreover, the BER performance is also affected by the symbol mapping probability
 (cf. Figure~$\ref{Result3}$).}
 
In Figure $\ref{Result3}$, we simulate the BER performance of Bob when QPSK is disguised as 16QAM in the single-antenna system. The symbol mapping follows the same rules defined in \emph{\textbf{Observation 1}}. As expected, the BER performance of the disguised signal is worse than that of the original low-order modulation due to the increased modulation order. However, the disguised signal consistently outperforms the target high-order modulation. This is because even when noise causes a symbol error, the misclassified symbol may still fall within the same mapping subset as the correct symbol, thus not affecting the final demapping outcome. 
Furthermore, we observe slight variations in the BER performance of Eve under different symbol mapping probabilities. This is due to the differing noise resilience of the constellation points in the mapped 16QAM diagram, which is influenced by their relative positions in the constellation space.

\begin{figure}[t]
\hspace{-10pt}
\includegraphics[width=9.5cm]{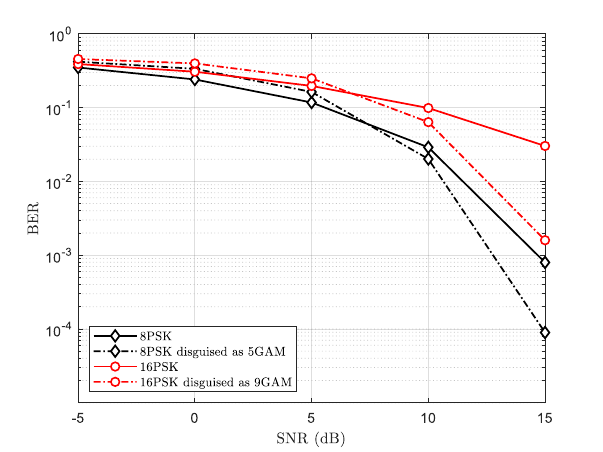}
\caption{The BER of Bob is evaluated w.r.t SNR for high-to-low-order confusion in the single-antenna system. In the simulation, 16PSK is disguised as 9GAM and 8PSK is disguised as 5GAM. 
Each frame consists of $10$ original modulation symbols.
The BER of conventional 16PSK and 8PSK is also simulated as benchmarks for comparison.}
\label{Result4}
\end{figure}

\emph{\textbf{Observation 3:}
In the single-antenna system, the proposed high-to-low MOC scheme provides noticeable BER performance gains at high SNR
 (cf. Figure~$\ref{Result4}$).}

In Figure $\ref{Result4}$, we apply STD to disguise 16PSK and 8PSK as 9GAM and 5GAM, respectively, with each frame consisting of $10$ original modulation symbols. Compared to the original high-order modulations, the system exhibits significantly improved BER performance at high SNR. This improvement is attributed to two factors.
First, the reduced modulation order enhances noise resilience.
Next, the proposed DP demodulator offers error correction capability by jointly decoding all the $10$ symbols in a frame for better performance. 
However, at low SNR, the BER performance of the disguised signals deteriorates compared to the original high-order modulations, due to severe noise-induced error propagation that impairs the effectiveness of the DP demodulator. 


\begin{figure}[t]
\hspace{-12pt}
\includegraphics[width=9.7cm]{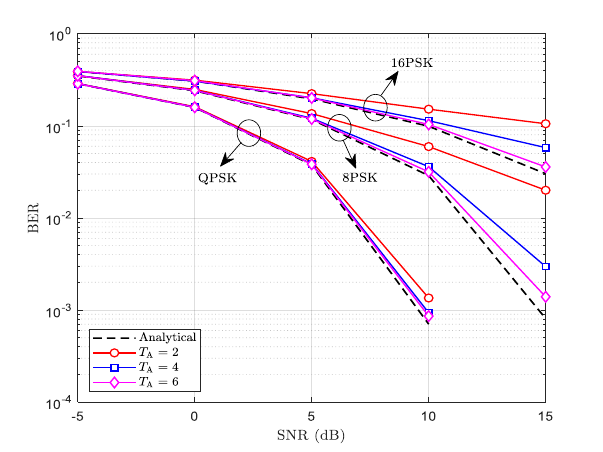}
\caption{The BER of Bob is evaluated w.r.t SNR for low-to-high-order confusion in the multi-antenna system. In the simulation, we use the Taylor-series-based confusion scheme and the nonlinear function is $f(s)=\arctan s$. 
The number of nonlinear terms in the Taylor expansion of $f(s)$ is set equal to  \( T_\text{A} - 1 \). }
\label{Result5}
\end{figure}
\begin{figure}[t]
\hspace{-12pt}
\includegraphics[width=9.68cm]{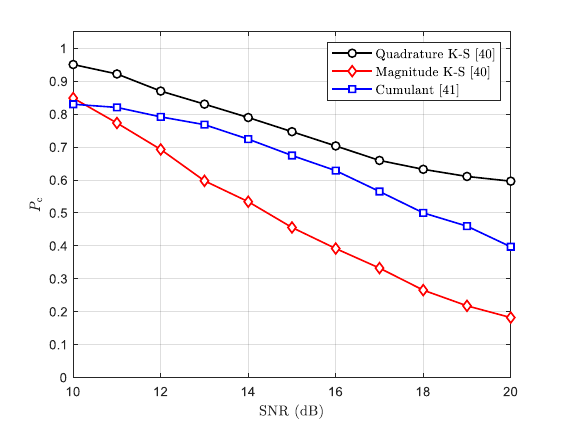}
\caption{The classification accuracy is evaluated w.r.t SNR for high-to-low-order confusion in the multi-antenna system. In the simulation, 16QAM is disguised as 4QAM.  We assume that Eve deploys three expert-knowledge-based classifiers, i.e., quadrature K-S classifier\cite{wang2010fast}, magnitude K-S classifier\cite{wang2010fast}, and cumulant-based  classifier\cite{swami2000hierarchical}. }
\label{Result6}
\end{figure}

\emph{\textbf{Observation 4:}
In the multi-antenna system, the Taylor-series-based confusion scheme only needs to eliminate the first $6$ nonlinear terms in the expansion of \( f(s) \) for Bob to successfully recover the original signal (cf. Figure~$\ref{Result5}$).}

In Figure~$\ref{Result5}$, we choose the nonlinear function \( f(s) = \arctan(s) \), and assign each antenna to eliminate one nonlinear term in its Taylor series. 
First, the results in different modulation schemes show that the BER performance of Bob improves as \( T_\text{A} \) increases. This is because a larger \( T_\text{A} \) allows for the cancellation of more nonlinear interference terms in the Taylor series, thereby enhancing the accuracy of signal recovery. 
Moreover, by comparing BER performance across different modulation schemes, we observe that for QPSK, the impact of \( T_\text{A} \) on BER is relatively minor compared to higher-order modulations. This is because nonlinear transforms such as \( f(s) = \arctan(s) \) have a limited effect on the sparse constellation of low-order modulations. 
However, for the high-order modulations, the constellation contains many symbols that are uniformly distributed, some nonlinear transform can severely destroy the original layout, requiring the elimination of a large number of interference terms to recover the original distribution. 
Furthermore, when \( T_\text{A} = 6 \), corresponding to the cancellation of $6$ nonlinear terms, the BER of Bob approaches the ideal case. Although increasing \( T_\text{A} \) beyond $6$ can further improve BER performance and enhance the confusion effect, it also results in higher computational complexity.

\emph{\textbf{Observation 5:}
In the multi-antenna system, even if Eve successfully extracts the signals from each antenna of Alice via blind source separation, the proposed security scheme can effectively prevent it from identifying the correct modulation format (cf. Figure~$\ref{Result6}$).}

In Figure $\ref{Result6}$, we simulate the anti-modulation-classification performance of the proposed high-to-low modulation confusion scheme in the multi-antenna system. 
The candidate modulation set includes 4QAM and 16QAM.
Alice uses $3$ antennas to disguise 16QAM as 4QAM following the CPD scheme illustrated in Figure $\ref{constellation}$. Eve is also equipped with $3$ antennas and applies fast independent component analysis \cite{hyvarinen2000independent} for blind source separation to recover the signals from each antenna of Alice, attempting to identify their modulation types individually. 
To more comprehensively evaluate the robustness of our scheme, and differing from \emph{\textbf{Observation 1}}, we assume that Eve employs three expert-knowledge-based classifiers, i.e., quadrature K-S classifier \cite{wang2010fast}, amplitude K-S classifier\cite{wang2010fast}, and cumulant-based classifier\cite{swami2000hierarchical}. 
The candidate modulation set is defined as 16QAM and 4QAM. Eve performs independent modulation classification on each separated signal and records the correct classification rate. The results show that the proposed scheme effectively prevents modulation classification across all the three classifiers. Similar to the deep-learning-based classifiers, the confusion performance improves as the SNR increases.

\section{Conclusion}
In this paper, we proposed a secure communication approach based on modulation order confusion, which ensures security by preventing Eve from correctly identifying the legitimate modulation format. 
For the single-antenna case, we introduced two modulation order confusion schemes, i.e., the symbol-random-mapping scheme, which disguises the low-order modulation as higher-order, and the symbol-time-diversity scheme, which disguises the high-order modulation as lower-order. 
For the multi-antenna system, we designed the transmit signals using either the series-expansion-based scheme or the constellation-path-design scheme.
Simulation results showed that the proposed schemes effectively defend against modulation classification by Eve in both the single-antenna and multi-antenna systems, including classifiers based on deep learning and expert knowledge. Furthermore, the anti-classification performance improves as the SNR increases. In the single-antenna system, the BER of the symbol-random-mapping scheme is influenced by the mapping probability, while the symbol-time-diversity scheme offers BER gains at high SNR. In the multi-antenna system, for the series-expansion-based scheme, it is recommended to use the first $6$ terms in the expansion to balance the computational complexity and the error performance. Furthermore, even if Eve successfully performs blind source separation and extracts signals from each antenna of Alice, it is unable to identify the true modulation format.
Compared with the existing secure communication and anti-modulation-classification techniques, the proposed approach does not require any knowledge of Eve. It also offers users the flexibility to disguise the original signal as a higher-order or lower-order modulation, thereby enabling a tunable balance between communication reliability and security.

\newpage
\appendix
\section*{Proof of Lemma 1}
The symbol distribution ${P_{{\text{ob}}}}({b_n})$ after modulation confusion with mapping 
probabilities $\boldsymbol{p}_m,~m\in {{\mathcal{I}}_{M}}$ is given by
\begin{equation}
{P_{{\text{ob}}}}({b_n}) = \sum\limits_{m = 1}^M {P({a_m})P({b_n}|{a_m})} 
\end{equation}
where 
\begin{align}
P({a_m}) &= \frac{1}{M}\\
P({b_n}|{a_m}) &= \left\{ \begin{gathered}
  p_i^{(m)},{\text{if }}{b_n}{\text{ is the }}i{\text{th symbol of }}{\mathcal{B}_m} \hfill \\
  0, \text{otherwise}. \hfill \\ 
\end{gathered}  \right.
\end{align}
The ideal symbol distribution ${P_{{\text{id}}}}({b_n})$ of the target modulation format to be confused is given by
\begin{equation}
{P_{{\text{id}}}}({b_n}) = \frac{1}{N}.
\end{equation}
The KL divergence between the distributions ${P_{{\text{ob}}}}$ and ${P_{{\text{id}}}}$ is given by
\begin{align}
{D_{{\text{KL}}}}({P_{{\text{ob}}}}||{P_{{\text{id}}}}) &= \sum\limits_{n = 1}^N {{P_{{\text{ob}}}}({b_n})\log \frac{{{P_{{\text{ob}}}}({b_n})}}{{{P_{{\text{id}}}}({b_n})}}} \\
& = \sum\limits_{m = 1}^M {\sum\limits_{i = 1}^K {\frac{1}{M}p_i^{(m)}\log N  \frac{1}{M}p_i^{(m)}} } \\
& = \sum\limits_{m = 1}^M {\sum\limits_{i = 1}^K {\frac{1}{M}p_i^{(m)}\big( {\log K + \log p_i^{(m)}} \big)} } \\
&  = \log K + \frac{1}{M}\sum\limits_{m = 1}^M {\sum\limits_{i = 1}^K {p_i^{(m)}\log p_i^{(m)}} } \\
& = \log K - \frac{1}{M}\sum\limits_{m = 1}^M {H({{\boldsymbol{p}}_m})}.
\end{align}
Then, let ${{{ \mathcal{\bar B}}}_m} = {\mathcal{B}}\setminus{{\mathcal{B}}_m},~m \in {{\mathcal{I}}_M}$ represent the complement of ${\mathcal{B}}_m$, denoted as
\begin{equation}
{{{ \mathcal{\bar B}}}_m} = \left\{ {\bar b_1^{(m)},\bar b_2^{(m)}, \ldots ,\bar b_{N - {K}}^{(m)}} \right\},~m \in {{\mathcal{I}}_M}.
\end{equation}
According to the union bound technique, $\vartheta $ can be approximately expressed as as
\begin{align}
\vartheta  &= \frac{1}{{M\log M}}\sum\limits_{m = 1}^M {\sum\limits_{i = 1}^{{K}} {p_i^{(m)}\sum\limits_{j = 1}^{N - {K}} {\rho _{i,j}^{(m)}p\left( {b_i^{(m)} \to \bar b_j^{(m)}} \right)} } } \\
&= \frac{1}{{M\log M}}\sum\limits_{m = 1}^M {\sum\limits_{i = 1}^{{K}} {p_i^{(m)}\sum\limits_{j = 1}^{N - {K}} {\rho _{i,j}^{(m)}Q\left( {\frac{{{\Delta _{i,j}^{(m)}}}}{{\sqrt 2 \sigma }}} \right)} } } 
\end{align}
where ${\rho _{i,j}^{(m)}}$ denotes the Hamming distance of the codewords between $b_i^{(m)}$ and $\bar b_j^{(m)}$, ${\Delta _{i,j}^{(m)}} = \big| {b_i^{(m)} - \bar b_j^{(m)}} \big|$, $Q(\cdot)$ is the Gaussian $Q$ function. 

\newpage
\bibliographystyle{IEEEtran}
\bibliography{ref}

@article{01nguyen2021security,
  title={Security and privacy for {6G}: A survey on prospective technologies and challenges},
  author={Nguyen, Van-Linh and Lin, Po-Ching and Cheng, Bo-Chao and Hwang, Ren-Hung and Lin, Ying-Dar},
  journal={IEEE Commun. Surveys Tuts.},
  volume={23},
  number={4},
  pages={2384--2428},
  year={2021},
  month={4th Quart.}
}

@article{02zou2016survey,
  title={A survey on wireless security: Technical challenges, recent advances, and future trends},
  author={Zou, Yulong and Zhu, Jia and Wang, Xianbin and Hanzo, Lajos},
  journal={Proc. IEEE},
  volume={104},
  number={9},
  pages={1727--1765},
  year={2016},
  month={Sep}
}

@article{03angueira2022survey,
  title={A survey of physical layer techniques for secure wireless communications in industry},
  author={Angueira, Pablo and Val, I{\~n}aki and Montalban, Jon and Seijo, {\'O}scar and Iradier, Eneko and Fontaneda, Pablo Sanz and Fanari, Lorenzo and Arriola, Aitor},
  journal={IEEE Commun. Surveys Tuts.},
  volume={24},
  number={2},
  pages={810--838},
  year={2022},
  month={2nd Quart.}
}

@article{04jameel2018comprehensive,
  title={A comprehensive survey on cooperative relaying and jamming strategies for physical layer security},
  author={Jameel, Furqan and Wyne, Shurjeel and Kaddoum, Georges and Duong, Trung Q},
  journal={IEEE Commun. Surveys Tuts.},
  volume={21},
  number={3},
  pages={2734--2771},
  year={2018},
  month={3rd Quart.}
}

@article{05bloch2008wireless,
  title={Wireless information-theoretic security},
  author={Bloch, Matthieu and Barros, Jo{\~a}o and Rodrigues, Miguel RD and McLaughlin, Steven W},
  journal={IEEE Trans. Inf. Theory},
  volume={54},
  number={6},
  pages={2515--2534},
  year={2008},
  month={Jun.}
}

@article{06goel2008guaranteeing,
  title={Guaranteeing secrecy using artificial noise},
  author={Goel, Satashu and Negi, Rohit},
  journal={IEEE Trans. Wireless Commun.},
  volume={7},
  number={6},
  pages={2180--2189},
  year={2008},
  month={Jun.}
}

@article{07huan1995likelihood,
  title={Likelihood methods for {MPSK} modulation classification},
  author={Huan, Chung-Yu and Polydoros, Andreas},
  journal={IEEE Trans. Commun.},
  volume={43},
  number={2/3/4},
  pages={1493--1504},
  year={1995},
  month={Feb.}
}

@INPROCEEDINGS{08yucek2004novel,
  title={A novel sub-optimum maximum-likelihood modulation classification algorithm for adaptive {OFDM} systems},
  author={Yucek, Tevfik and Arslan, H{\"u}seyin},
  booktitle={Proc. 2004 IEEE Wireless Commun. $\emph \&$ Netw. Conf. (WCNC)},
  volume={2},
  pages={739--744},
  year={2004},
  month={Jul.}
}

@article{09zhu2018likelihood,
  title={A likelihood-based algorithm for blind identification of {QAM} and {PSK} signals},
  author={Zhu, Daimei and Mathews, V John and Detienne, David H},
  journal={IEEE Trans. Wireless Commun.},
  volume={17},
  number={5},
  pages={3417--3430},
  year={2018},
  month={May}
}

@article{10wei2000maximum,
  title={Maximum-likelihood classification for digital amplitude-phase modulations},
  author={Wei, Wen and Mendel, Jerry M},
  journal={IEEE Trans. Commun.},
  volume={48},
  number={2},
  pages={189--193},
  year={2000},
  month={Feb.}
}

@article{11swami2000hierarchical,
  title={Hierarchical digital modulation classification using cumulants},
  author={Swami, Ananthram and Sadler, Brian M},
  journal={IEEE Trans. Commun.},
  volume={48},
  number={3},
  pages={416--429},
  year={2000},
  month={Mar.}
}

@article{12wu2008novel,
  title={Novel automatic modulation classification using cumulant features for communications via multipath channels},
  author={Wu, Hsiao-Chun and Saquib, Mohammad and Yun, Zhifeng},
  journal={IEEE Trans. Wireless Commun.},
  volume={7},
  number={8},
  pages={3098--3105},
  year={2008},
  month={Aug.}
}

@article{15dobre2010cyclostationarity,
  title={Cyclostationarity-based modulation classification of linear digital modulations in flat fading channels},
  author={Dobre, Octavia A and Abdi, Ali and Bar-Ness, Yeheskel and Su, Wei},
  journal={Wireless Personal Commun.},
  volume={54},
  number={4},
  pages={699--717},
  year={2010},
}

@inproceedings{14dobre2004robust,
  title={Robust {QAM} modulation classification algorithm using cyclic cumulants},
  author={Dobre, Octavia A and Bar-Ness, Yeheskel and Su, Wei},
  booktitle={Proc. 2004 IEEE Wireless Commun. $\emph \&$ Netw. Conf. (WCNC)},
  volume={2},
  pages={745--748},
  year={2004},
  month={Mar.}
}

@article{13chang2015cumulants,
  title={Cumulants-based modulation classification technique in multipath fading channels},
  author={Chang, Dah-Chung and Shih, Po-Kuan},
  journal={IET Commun.},
  volume={9},
  number={6},
  pages={828--835},
  year={2015},
  month={Apr.}
}

@inproceedings{16park2008automatic,
  title={Automatic modulation recognition of digital signals using wavelet features and {SVM}},
  author={Park, Cheol-Sun and Choi, Jun-Ho and Nah, Sun-Phil and Jang, Won and Kim, Dae Young},
  booktitle={Proc. Int. conf. Adv. Commun. Technol.},
  volume={1},
  pages={387--390},
  year={2008},
}

@article{17o2017introduction,
  title={An introduction to deep learning for the physical layer},
  author={O’Shea, Timothy James and Hoydis, Jakob},
  journal={IEEE Trans. Cognit. Commun. Netw.},
  volume={3},
  number={4},
  pages={563--575},
  year={2017},
  month={Dec.}
}

@article{22njoku2021cgdnet,
  title={{CGDNet}: Efficient hybrid deep learning model for robust automatic modulation recognition},
  author={Njoku, Judith Nkechinyere and Morocho-Cayamcela, Manuel Eugenio and Lim, Wansu},
  journal={IEEE Netw. Lett.},
  volume={3},
  number={2},
  pages={47--51},
  year={2021},
  month={Jun}
}

@article{24choqueuse2009blind,
  title={Blind modulation recognition for {MIMO} systems},
  author={Choqueuse, Vincent and Azou, Stephane and Yao, Koffi and Collin, Ludovic and Burel, Gilles},
  journal={MTA Review},
  volume={19},
  number={2},
  pages={183--196},
  year={2009},
  month={Jun.}
}

@inproceedings{25zhu2014blind,
  title={Blind modulation classification for {MIMO} systems using expectation maximization},
  author={Zhu, Zhechen and Nandi, Asoke K},
  booktitle={Proc. IEEE MILCOM},
  pages={754--759},
  year={Baltimore, MD, USA, 2014},
}

@article{26liu2017blind,
  title={Blind modulation classification algorithm based on machine learning for spatially correlated {MIMO} system},
  author={Liu, Xiaokai and Zhao, Chenglin and Wang, Pengbiao and Zhang, Yang and Yang, Tianpu},
  journal={IET Commun.},
  volume={11},
  number={7},
  pages={1000--1007},
  year={2017},
  month={May}
}

@article{27hassan2011blind,
  title={Blind digital modulation identification for spatially-correlated {MIMO} systems},
  author={Hassan, Kais and Dayoub, Iyad and Hamouda, Walaa and Nzeza, Crepin Nsiala and Berbineau, Marion},
  journal={IEEE Trans. Wireless Commun.},
  volume={11},
  number={2},
  pages={683--693},
  year={2011},
  month={Feb.}
}

@article{28althunibat2017physical,
  title={A physical-layer security scheme by phase-based adaptive modulation},
  author={Althunibat, Saud and Sucasas, Victor and Rodriguez, Jonathan},
  journal={IEEE Trans. Veh. Technol.},
  volume={66},
  number={11},
  pages={9931--9942},
  year={2017},
  month={Nov}
}

@article{29bang2020secure,
  title={Secure modulation based on constellation mapping obfuscation in {OFDM} based {TDD} systems},
  author={Bang, Inkyu and Kim, Taehoon},
  journal={IEEE Access},
  volume={8},
  pages={197644--197653},
  year={2020},
  month={Nov.}
}

@article{34sadeghi2018adversarial,
  title={Adversarial attacks on deep-learning based radio signal classification},
  author={Sadeghi, Meysam and Larsson, Erik G},
  journal={IEEE Wireless Commun. Lett.},
  volume={8},
  number={1},
  pages={213--216},
  year={2018},
  month={Feb.}
}

@article{35lin2020adversarial,
  title={Adversarial attacks in modulation recognition with convolutional neural networks},
  author={Lin, Yun and Zhao, Haojun and Ma, Xuefei and Tu, Ya and Wang, Meiyu},
  journal={IEEE Trans. Rel.},
  volume={70},
  number={1},
  pages={389--401},
  year={2021},
  month={Mar.}
}

@article{36kim2021channel,
  title={Channel-aware adversarial attacks against deep learning-based wireless signal classifiers},
  author={Kim, Brian and Sagduyu, Yalin E and Davaslioglu, Kemal and Erpek, Tugba and Ulukus, Sennur},
  journal={IEEE Trans. Wireless Commun.},
  volume={21},
  number={6},
  pages={3868--3880},
  year={2022},
  month={Jun.}
}

@article{37he2023anti,
  title={Anti-modulation-classification Transmitter Design Against Deep Learning Approaches},
  author={He, Boxiang and Wang, Fanggang},
  journal={IEEE Trans. Wireless Commun.},
  volume={23},
  number={7},
  pages={6808--6823},
  year={2024},
  month={Jul.}
}

@article{larsson2017golden,
  title={Golden angle modulation},
  author={Larsson, Peter},
  journal={IEEE Wireless Commun. Lett.},
  volume={7},
  number={1},
  pages={98--101},
  year={2017},
  month={Sep.}
}

@misc{grant2009cvx,
  title={{CVX}: Matlab software for disciplined convex programming},
  author={Grant, Michael and Boyd, Stephen and Ye, Yinyu},
  year={2009}
}

@article{zheng2018ultra,
  title={Ultra-wideband side-lobe level suppression using amplitude-adjustable metasurfaces},
  author={Zheng and others},
  journal={J. Phys. D: Appl. Phys.},
  volume={52},
  number={6},
  pages={065102},
  year={2018},
  month={Dec.}
}

@article{9luo2021reconfigurable,
  title={Reconfigurable intelligent surface: Reflection design against passive eavesdropping},
  author={Luo, Junshan and Wang, Fanggang and Wang, Shilian and Wang, Hao and Wang, Dong},
  journal={IEEE Trans. Wireless Commun.},
  volume={20},
  number={5},
  pages={3350--3364},
  year={2021},
  month={May}
}

@article{he2019cascaded,
  title={Cascaded channel estimation for large intelligent metasurface assisted massive {MIMO}},
  author={He, Zhen-Qing and Yuan, Xiaojun},
  journal={IEEE Wireless Commun. Lett.},
  volume={9},
  number={2},
  pages={210--214},
  year={2020},
  month={Feb.}
}

@article{yan2018secret,
  title={Secret channel training to enhance physical layer security with a full-duplex receiver},
  author={Yan, Shihao and Zhou, Xiangyun and Yang, Nan and Abhayapala, Thushara D and Swindlehurst, A Lee},
  journal={IEEE Trans. Inf. Forensics Security},
  volume={13},
  number={11},
  pages={2788--2800},
  year={2018},
  month={Nov.}
}

@article{VGGo2018over,
  title={Over-the-air deep learning based radio signal classification},
  author={O’Shea, Timothy James and Roy, Tamoghna and Clancy, T Charles},
  journal={IEEE J. Sel. Top. Signal Process.},
  volume={12},
  number={1},
  pages={168--179},
  year={2018},
  month={Jan.}
}

@article{WSMFqi2020automatic,
  title={Automatic modulation classification based on deep residual networks with multimodal information},
  author={Qi, Peihan and Zhou, Xiaoyu and Zheng, Shilian and Li, Zan},
  journal={IEEE Trans. Cogn. Commun. Netw.},
  volume={7},
  number={1},
  pages={21--33},
  year={2021},
  month={March}
}

@article{SCGNettunze2020sparsely,
  title={Sparsely connected {CNN} for efficient automatic modulation recognition},
  author={Tunze, Godwin Brown and Huynh-The, Thien and Lee, Jae-Min and Kim, Dong-Seong},
  journal={IEEE Trans. Veh. Technol.},
  volume={69},
  number={12},
  pages={15557--15568},
  year={2020},
  month={Dec.}
}

@inproceedings{ChainNethuynh2020chain,
  title={Chain-Net: Learning deep model for modulation classification under synthetic channel impairment},
  author={Huynh-The, Thien and Doan, Van-Sang and Hua, Cam-Hao and Pham, Quoc-Viet and Kim, Dong-Seong},
  booktitle={Proc. IEEE Global Commun. Conf. (GLOBECOM)},
  pages={1--6},
  year={2020},
  month={Dec.}
}

@article{wang2025secure,
  title={Secure Communication via Nonlinear Transmit and Collaborative Relaying},
  author={Wang, Jingyi and Wang, Fanggang and He, Boxiang and Luo, Junshan},
  journal={IEEE Trans. Commun.},
  year={2025},
  publisher={IEEE}
}

@article{hyvarinen2000independent,
  title={Independent component analysis: Algorithms and applications},
  author={Hyv{\"a}rinen, Aapo and Oja, Erkki},
  journal={Neural Netw.},
  volume={13},
  number={4-5},
  pages={411--430},
  year={2000},
  month={June}
}

@article{wang2010fast,
  title={Fast and robust modulation classification via Kolmogorov-Smirnov test},
  author={Wang, Fanggang and Wang, Xiaodong},
  journal={IEEE Trans. Commun.},
  volume={58},
  number={8},
  pages={2324--2332},
  year={2010},
  month={Aug.}
}

@article{swami2000hierarchical,
  title={Hierarchical digital modulation classification using cumulants},
  author={Swami, Ananthram and Sadler, Brian M},
  journal={IEEE Trans. Commun.},
  volume={48},
  number={3},
  pages={416--429},
  year={2000},
  month={Mar}
}

\end{document}